\documentclass[a4paper,11pt]{article}
\pdfoutput=1 

\usepackage{jcappub} 

\usepackage[T1]{fontenc} 

\usepackage{graphicx,natbib,amssymb,amsmath,stmaryrd,makecell}
\usepackage{txfonts}
\usepackage{xcolor}
\usepackage{hyperref}
\usepackage{xspace,dsfont}
\usepackage{caption, lipsum, comment}

\graphicspath{{Figures/}}

\newcommand{\expf}[1]{{{\rm e}^{#1}}}

\newcommand{\vgh}{{\hat{\boldsymbol\gamma}}}

\newcommand{\vb}{{\boldsymbol{\beta}}}
\newcommand{\vbh}{{\boldsymbol{\hat{\beta}}}}

\newcommand{\beq}{\begin{equation}}   %

\newcommand{\eeq}{\end{equation}}   %

\newcommand{\beqa}{\begin{eqnarray}}   %

\newcommand{\eeqa}{\end{eqnarray}}   %

\newcommand{\bealf}[1]{\begin{align} #1 \end{align}}

\newcommand{\beal}{\begin{align}}
\newcommand{\enal}{\end{align}}

\newcommand{\bspl}{\begin{split}}

\newcommand{\espl}{\end{split}}

\newcommand{\bsub}{\begin{subequations}}

\newcommand{\esub}{\end{subequations}}

\newcommand{\bmulti}{\begin{multline}}   %

\newcommand{\beqm}{\begin{mathletters}}   %

\newcommand{\eeqm}{\end{mathletters}}   %

\newcommand{\The}{\theta_{\rm e}}

\newcommand{\oDnu}{{\mathcal{\hat{D}}_{\nu}}}

\newcommand{\Sll}[2]{\hat{\mathcal{S}}^{#1}_{#2}}

\newcommand{\oOnu}{{\mathcal{\hat{O}}_{\nu}}}

\usepackage{comment}

\newcommand{\Bb}[4]{{^{#1}_{#2}}\mathcal{\hat{B}}^{#3}_{#4}}

\newcommand{\Dbo}[3]{{{^{#1}}\mathcal{\hat{D}}^{#3}_{#2}}}

\title{Simplified treatment of kinematic corrections to the SZ effect using the boost operator approach}

\author[a]{Alex Hoey,}
\author[a]{Jacob Long}
\author[a]{and Jens Chluba}

\emailAdd{alex.hoey@student.manchester.ac.uk}
\emailAdd{jacob.long@student.manchester.ac.uk}
\emailAdd{jens.chluba@manchester.ac.uk}

\affiliation[a]{Jodrell Bank Centre for Astrophysics, School of Physics and Astronomy, The University of Manchester, Oxford Road, Manchester, M13 9PL, U.K.}

\date{Feb 2026}

\begin{document}

\abstract{The Sunyaev-Zeldovich (SZ) effects provide a potent cosmological probe of the large-scale structure in the universe. Here, we present a simplified derivation of the kinematic corrections to the relativistic thermal SZ effect using the boost operator approach. By first performing the thermal average inside the moving electron cloud frame, the thermal and peculiar motion contributions can be naturally separated, leading to a significant simplification of the scattering calculation. The angular dependence of the problem is resummed into the pre-computed Doppler operators avoiding the otherwise cumbersome many-dimensional angular integrals required using conventional approaches. 
We provide expressions for the relativistic SZ signal, exact to all orders in the electron temperature $\The$ and the peculiar velocity $\beta_{\text{p}}$ of a given cluster. These reproduce well-known results for the relativistic SZ effects and extend the description to higher orders in the cluster's speed. We also derive a closed-form expression for the thermally-averaged thermal SZ scattering operator by studying the underlying symmetries of the Doppler operators fundamental to the formalism. Through these results, we also demonstrate how the boost operator approach gives clarity to the physical description of the problem at hand, promising to be useful to a wide range of astrophysical problems.}

\maketitle
\flushbottom
\vspace{5mm}

\section{Introduction}
\label{sec:Intro}
The scattering of cosmic microwave background (CMB) photons by free electrons gives rise to the Sunyaev--Zeldovich (SZ) effects. Random electron motions in the hot intracluster medium generate the thermal SZ (tSZ) effect, whose amplitude traces the integrated electron pressure, while the coherent motion of the scattering medium relative to the CMB produces the kinematic SZ (kSZ) effect, which probes its line-of-sight momentum \citep{Zeldovich1969,Sunyaev1970,Sunyaev1980}. The absence of cosmological surface-brightness dimming makes the SZ effects particularly valuable for detecting and characterising galaxy clusters over a wide range of redshifts. They have consequently become important probes of cluster astrophysics, the distribution of ionised matter and the growth of large-scale structure \citep[e.g.][]{Birkinshaw1999,Carlstrom2002,Mroczkowski2019}.

The intracluster medium of massive clusters commonly reaches temperatures of several keV, with shock-heated regions in merging systems extending to substantially higher temperatures \citep[e.g.,][]{Refregier2000,Kay2008, Lee2020SZ, Lee2022SZ}. The electron population can therefore be mildly relativistic, and the usual non-relativistic description of the tSZ effect becomes insufficient. Relativistic corrections modify the amplitude and spectral shape of the distortion, shift the tSZ null and introduce sensitivity to the temperature structure of the scattering medium \citep{Wright1979,Fabbri1981,Challinor1998,Itoh98, Chluba2012SZpack, Chluba2012moments}. These temperature corrections are now close to being detected both statistically and in individual clusters, demonstrating their relevance for the interpretation of increasingly precise multifrequency SZ observations \citep{Hansen2002, Zemcov2012, Hurier2016,Erler2017,Remazeilles2019CellrSZ,
Butler2022rSZ, Remazeilles2025MNRAS, Coulton2026rSZ}.

Relativistic corrections are also important for the kSZ effect. In the non-relativistic limit, the kSZ signal manifests as a blackbody temperature shift proportional to the Thomson optical depth and the line-of-sight peculiar velocity, $\Delta T/T_{\rm CMB}\simeq-\tau\beta_{\rm p}\mu_{\rm p}$. Here, $\beta_{\rm p}$ is the cluster's speed and $\mu_{\rm p}$ the direction cosine with respect to the line-of-sight. At finite electron temperature, the SZ signal acquires thermal--kinematic corrections proportional to $\beta_{\rm p}\The^n$, where $\The=k_{\rm B}T_{\rm e}/m_{\rm e}c^2$, together with higher-order terms in the bulk velocity \citep{Sazonov1998,Nozawa1998SZ,Nozawa2006} and observer motion \citep{Chluba2005b,Nozawa2005, Chluba2012SZpack}. These contributions become particularly relevant in hot, rapidly moving merging systems. For example, MACS~J0717.5+3745 contains shock-heated gas with temperatures of order $20\,{\rm keV}$ and a subcluster moving at a line-of-sight velocity of approximately $3000\,{\rm km\,s^{-1}}$, towards which a spatially resolved kSZ signal has been measured \citep{Mroczkowski2012,Sayers_2013,refId0}. Simulations likewise indicate that the kSZ signal can become comparable to, or locally dominate over, the tSZ signal in suitable projections of merging clusters \citep{Ruan2013mergerSZ}. Accurate relativistic modelling is therefore required if future high precision SZ observations of such systems are to provide unbiased measurements of their temperatures and velocities. Future facilities such as AtLAST \citep{DiMascolo2025AtLAST} and CMB-HD \citep{Sehgal2019CMBHD} are being developed to provide sensitive measurements of the full SZ spectrum, improving the separation of its thermal, kinematic and relativistic contributions.

Several approaches have been developed to calculate the relativistic thermal and kinematic SZ signals, including direct evaluation of the Boltzmann collision term and expansions in $\The$ and $\beta_{\rm p}$ \citep[e.g.][]{Challinor1998,Itoh98,Sazonov1998,Nozawa1998SZ,Chluba2012SZpack}. These methods provide accurate predictions and form the basis of numerical tools such as {\tt SZpack} \citep{Chluba2012SZpack, Chluba2012moments}. However, when the calculation is carried out in the CMB rest frame, the electron distribution simultaneously contains the random thermal motion and the bulk peculiar velocity of the scattering medium. The corresponding relativistic Maxwellian is anisotropic \citep[e.g.,][]{Lee2024SZpack}, coupling the thermal and kinematic variables within the momentum and angular integrations. Although the final result can be organised as an expansion in $\The$ and $\beta_{\rm p}$, the two physical effects remain intertwined during the calculation, making higher-order corrections increasingly cumbersome.

In this work, we present a different organisation of the calculation using the recently developed boost operator formalism \citep{ChlubaBO25}, which was already applied to SZ calculations \citep{Chluba2026SZ,Rosenberg2025prSZ} and the Kompaneets equation \citep{hoey2026derivationkompaneetsequationusing}. In our problem, we first transform the incident radiation into the rest frame of the electron cloud, where the electron distribution is isotropic and the thermal average can be performed independently of the cloud's peculiar velocity. The resulting thermally-averaged scattering operator is then transformed back to the CMB frame using the boost and Doppler operators.
%
This ordering cleanly factorises the thermal and bulk-motion dependences: the thermal physics is contained in a set of scattering operators that depend only on $\The$, while the peculiar motion is encoded by operators depending only on $\beta_{\rm p}$. The angular structure is resummed into the Doppler operators, avoiding the repeated multidimensional angular integrations that arise in conventional treatments.

The resulting SZ operator is {\it exact} in both the electron temperature and the peculiar velocity before either expansion is performed. The thermal and kinematic contributions may therefore be expanded independently to whatever order is required. We use this formulation to recover the known relativistic kSZ corrections, including terms through $\mathcal{O}(\beta_{\rm p}^{3})$ and the leading thermal--kinematic corrections, and to clarify the operator structure from which they arise. As part of our derivations, we also obtain a closed-form expansion for the thermally averaged monopole tSZ scattering operator.

The remainder of the paper is organised as follows. Section~\ref{K and B} briefly introduces the boost and Doppler operators used throughout the calculation.
In Section~\ref{kinematic sz main derivation}, we derive the relativistic SZ operator and compare its expansions with previous results. Section~\ref{app: closed form scattering} investigates the structure of the Doppler operators and derives the closed-form monopole tSZ scattering operator. We summarise our conclusions in Section~\ref{sec:Conc}.

\section{Introduction to the boost operator}
\label{K and B}
\noindent 
As might be expected, fundamental to the boost operator formalism is the boost operator \citep{ChlubaBO25}. In this section, we will briefly introduce the boost operator, along with some of its important properties required in the derivation. A more in depth discussion of the boost operator and aberration kernel is given in \citep{Dai2014,ChlubaBO25,hoey2026derivationkompaneetsequationusing}.

Plainly, the boost operator describes how the multipole coefficients of frequency dependent observables, that are described on a sphere, transform between frames. For some observable $X$, with Doppler weight $d$,\footnote{An observable with Doppler weight, $d$, will transform as $X'(\boldsymbol{\hat{n}}')
= 
\left(\nu'/\nu\right)^d\,
X(\boldsymbol{\hat{n}}\,[\boldsymbol{\hat{n}}']).
$} and multipole coefficients $X_{\ell' m'}$ in a rest frame S, and $X'_{\ell m}$ in a frame S' (with a general velocity $\boldsymbol{\beta}=\boldsymbol{\varv}/c$ relative to S), the transformation law is given by \citep{ChlubaBO25}
\bealf{
\label{app:Boost operator}
X_{\ell m}'(\nu)
&= 
\sum_{\ell' m'} {}^{d}_{}\mathcal{\hat{B}}^{mm'}_{\ell \ell'}(\nu, \boldsymbol{\beta}) X_{\ell'm'}(\nu). 
}
Here, ${}^{d}_{}\mathcal{\hat{B}}^{mm'}_{\ell \ell'}(\nu, \boldsymbol{\beta})$ is the boost operator labelled by the multipole numbers $\ell,\ell',m,m'$, and $\nu$ is the frequency. For boosts along the $z$-axis, the boost operator can be explicitly defined by 
\bealf{
\label{Boost op full 2}
{}^{d}_{}\mathcal{\hat{B}}^{m}_{\ell \ell'}(\nu, \beta)
&=\int \text{d}\vgh'\,\frac{{}_{}Y_{\ell m}^*(\vgh'){}_{}Y_{\ell' m'}(\vgh)}{[\gamma\,(1+\beta\mu')]^{\,d+\oOnu}},
}
where $Y_{\ell m}$ are spherical harmonic functions, $\vgh'$ and $\vgh$ are the respective directions on the sky in the rest and observer frames, $\mu'=\boldsymbol{\beta}\cdot \vgh'$ is the direction cosine, $\gamma = 1/\sqrt{1-\beta^2}$ is the Lorentz factor, and $\oOnu=-\nu\partial_\nu$ is the \textit{energy shift generator}.
In general, observables can also have a spin weight $s$, however we only consider observables with $s=0$, so we will neglect it in the notation. 

The boost operator for a general direction $\hat{\boldsymbol{\beta}}$ can be expressed in terms of the simpler $z$-aligned boost operator through a rotation. For a boost along the $z$-direction, there is no azimuthal ($m$) mixing, hence the boost operators are more readily calculable. This rotation is given by \citep[e.g., see][]{hoey2026derivationkompaneetsequationusing}
\bealf{
\label{eq:General boost definition}
\Bb{d}{}{mm'}{\ell\ell'}(\nu',\boldsymbol{\beta})=\sum_{m_1}D_{m m_1}^{\ell}\Bb{d}{}{m_1}{\ell\ell'}(\nu',\beta)[D_{m' m_1}^{\ell'}]^*,
}
Where $D_{m' m_1}^{\ell'}=D_{m' m_1}^{\ell'}(\phi, \theta, \psi)$ and $[D_{m m_1}^{\ell}]^*=[D_{m m_1}^{\ell}(\phi, \theta, \psi)]^*$ are Wigner-D functions and $\Bb{d}{}{m_1}{\ell'\ell}(\nu',\beta)$ is the $z$-aligned boost operator. Furthermore, the boost operator can be written in terms of the simpler aberration kernel \citep{Challinor2002, Dai2014} as \citep[see][for proof]{ChlubaBO25}
\bealf{
\label{eq:boost_operator}
{}^{d}\hat{\mathcal{B}}^{mm'}_{\ell \ell'}(\nu,\boldsymbol{\beta})
\equiv {}^{d+\oOnu}\hat{\mathcal{K}}^{mm'}_{\ell \ell'}(-\boldsymbol{\beta}),
}
where ${}^{d+\oOnu}\hat{\mathcal{K}}^{mm'}_{\ell \ell'}(-\boldsymbol{\beta})$ is the aberration kernel.\footnote{The aberration kernel is the equivalent of the boost operator for frequency-independent observables.}
The aberration kernel is well studied, and has a number of symmetries and recursion relations which can, in general, be trivially extended to the boost operator \citep[see][]{Chluba2011ab,
Dai2014,ChlubaBO25}. Consequently, the boost operator elements can be simply obtained using {\tt Mathematica}. The simplest example, and one we use extensively, is
\bealf{
\label{eq: Kernel example}
{}^{d}_{}\hat{\mathcal{B}}_{0 0}^{0}\,(\nu, \beta)=
{}^{d+\oOnu}_{}\hat{\mathcal{K}}_{0 0}^{0}\,(-\beta)
=\frac{\left(\gamma-p\right)^{1-d-\oOnu}-\left(\gamma+p\right)^{1-d-\oOnu}}{2\left(d+\oOnu-1\right)p},
}
where $p=\beta\gamma$ is the dimensionless electron momentum. We emphasise here that both the aberration kernel and the boost operator are exact functions of $p$, and should be thought of as 'special functions' of the frame transformation problem. 

It is helpful to define the {\it Doppler operator}, which is constructed from a pair of boost operators and frequently appears when describing radiative transfer problems when using the boost operator formalism. The Doppler operator is defined by 
\bealf{
\label{app:D}
^d\hat{\mathcal{D}}^m_{\ell\ell'\ell''}(\nu,\beta)\equiv\frac{1}{\gamma}{}^{d}\hat{\mathcal{B}}^m_{\ell \ell'}(\nu,-\beta)\,{}^{0}\hat{\mathcal{B}}^{m}_{\ell' \ell''}(\nu,\beta),
}
and describes how an input multipole $\ell''$ is transformed to an output multipole $\ell$, through an intermediate multipole $\ell'$ for boosts along the $z$-axis.

\vspace{-3mm}
\section{Derivation of the kinematic SZ effect}
\label{kinematic sz main derivation}
We begin by considering the scattering in the rest frame of an electron cloud defined as the frame in which the electron distribution is isotropic. The spectral distortion in this frame is similar to that of the thermal SZ only, but with an anisotropic incoming photon distribution. The spectral distortion due to the thermal SZ effect was calculated using the boost operator approach in \cite{Chluba2026SZ, Rosenberg2025prSZ}, and can be extended to a general photon distribution, $n'(\nu',\vgh')=\sum_{\ell m}n'_{\ell m}(\nu')\,Y'_{\ell m}(\vgh')$, by generalising the thermal SZ scattering operator to any multipole $\ell$. Here, the prime denotes quantities defined in the cloud frame (while unprimed quantities are defined in the lab frame). For the SZ signal in the cloud frame, we then have
\bsub
\bealf{
\label{eq:thermal correction Sll}
\Delta n_{\textrm{th}}'(\nu',\vgh')
&= \tau' \sum_{\ell m} Y_{\ell m}(\vgh') \,\Sll{\textrm{th}}{\ell} (\nu',\The)\,n_{\ell m}'(\nu'),
\\
\label{eq:Sll}
\Sll{\textrm{th}}{\ell}(\nu', \The) &= \int_0^{\infty} p^2 f(\gamma) \,\textrm{d}p \,\left[\frac{\Dbo{}{\ell 0 \ell}{}(\nu', \beta)}{2\ell + 1}+ \frac{1}{10}\frac{\Dbo{}{\ell 2 \ell}{}(\nu', \beta)}{2\ell+1} - 1\right],
\\
\label{relativistic maxwellian}
f(\gamma) &= \frac{\exp(-\gamma/\The)}{\The K_2(1/\The)}, \qquad \Dbo{}{\ell \ell' \ell''}{}(\nu', \beta)=\sum_{m'}\,\Dbo{-1}{\ell \ell' \ell''}{m'}(\nu', \beta)
,
}
\esub
where $n_{\ell m}'(\nu')$ are the photon distribution multipole coefficients, $\tau'$ is the Thomson scattering optical depth, $\Sll{\textrm{th}}{\ell}(\nu', \The)$ is the thermally averaged SZ operator, $f(\gamma)$ is a relativistic Maxwellian  \cite[see Eq.~(16) of][with $i=j=0$]{Rosenberg2025prSZ}, and $K_n$ is the modified Bessel function of the second kind. For convenience, we have defined the $m$-averaged Doppler operators $\Dbo{}{\ell \ell' \ell''}{}(\nu', \beta)$, dropping the Doppler-weight in the notation.\footnote{When considering recoil corrections, we observe a lowering of the Doppler weight by $-1$ for each successive order in recoil \citep[see][]{hoey2026derivationkompaneetsequationusing}, meaning that in that case the notation would need to be more general. We also note that the sum over $m'$ is determined by the minimal value of the three angular momentum quantum numbers.} 
Note that Eq.~\eqref{eq:thermal correction Sll} is valid in the case where the scattering medium is optically thin and multiple scatterings \citep[e.g.,][]{Chluba2014mSZI, Chluba2014mSZII} can be neglected. Additionally, Eq.~\eqref{eq:Sll} is already averaged over all boost directions $\text{d}\vbh$, which is trivially achieved due to the isotropy of the electron distribution in the cloud's rest frame \citep[see Eq.~(15) and (16) of][]{Rosenberg2025prSZ}. In practice, the thermal average can be carried out via an expansion of the Doppler operators in orders of $p$, and applying momentum moments \citep[see][]{CSpack2019}. 

Relative to the lab frame, the electron cloud moves with a peculiar velocity $\boldsymbol{\beta}_{\rm p}$ in an arbitrary direction; to express the correction in terms of lab frame quantities, we must apply a boost along the direction $\hat{\boldsymbol{\beta}}_{\rm p}$. The transformation of $n_{\ell m}'(\nu')$ is given by
\bealf{
\label{eq:sph harm coeff transform}
n_{\ell m}'(\nu') = \sum_{\ell' m'}\Bb{0}{}{mm'}{\ell\ell'}(\nu',\boldsymbol{\beta}_{\rm p})\,n_{\ell' m'}(\nu'),
}
 since the photon occupation number has $d=0$. This expression can be directly inserted into Eq.~\eqref{eq:thermal correction Sll}, giving the cloud frame signal in terms of the lab frame spherical harmonic coefficients of the photon occupation number, $n(\nu, \vgh)$.
 
 After applying the scattering operator, we continue by transforming back into the lab frame through a boost in the $-\hat{\boldsymbol{\beta}}_{\rm p}$ direction, for which we must also consider the transformation of the optical depth. We know \citep[e.g.,][]{Chluba2012SZpack, ChlubaBO25}\footnote{Since we assume $\beta_{\rm p}$ is constant along the line of sight.}
\bealf{
\label{eq:tau transform}
\text{d}\tau'&=(1-\beta_{\rm p}\mu_{\text{p}})\,\text{d}\tau=\Bigg(\frac{\nu'}{\nu}\Bigg)\,\frac{\text{d}\tau}{\gamma}
\Rightarrow
\tau'=\Bigg(\frac{\nu'}{\nu}\Bigg)\,\frac{\tau}{\gamma},
}
where $\mu_{\text{p}}=\vgh\cdot\vbh_{\text{p}}$ is the direction cosine of the incoming photon with the peculiar velocity. As a consequence of the $\tau$ transformation, $\Delta n'_{\textrm{th}}$ transforms back to the lab frame with a Doppler weight reduced by one ($d=-1$) and a factor of $1/\gamma$ \citep[see e.g.,][]{Rosenberg2025prSZ,Chluba2026SZ,hoey2026derivationkompaneetsequationusing}. Additionally, we can make the simple replacement $\Sll{\textrm{th}}{\ell} (\nu',\The)\rightarrow \Sll{\textrm{th}}{\ell} (\nu,\The)$ since it is constructed from the dimensionless energy shift generator $\oOnu$ only. Having followed the procedure set out above, applying Eq.~\eqref{eq:sph harm coeff transform} and boosting back into the lab frame, the multipole coefficients of Eq.~\eqref{eq:thermal correction Sll} become
\bealf{
\label{eq:lab frame correction sll}
\Delta n^{\textrm{th}}_{\ell m}(\nu)
=\frac{\tau}{\gamma_{\rm p}}\sum_{\ell' m'}\sum_{\ell'' m''} \Bb{-1}{}{m m'}{\ell\ell'}(\nu,-\boldsymbol{\beta}_{\rm p})\,\Sll{\textrm{th}}{\ell'}(\nu,\The)\,\Bb{0}{}{m' m''}{\ell'\ell''}(\nu,\boldsymbol{\beta}_{\rm p})\,\,n_{\ell'' m''}
(\nu)
.
}
We note here that $\tau$ is the lab-frame optical depth which includes kinematic corrections, meaning $\tau=\tau'/(1-\beta_{\rm p}\mu_{\rm p})$ \citep{Chluba2012SZpack}.

To express the above equation in terms of the simpler $z$-aligned boost operators, we apply Eq.~\eqref{eq:General boost definition} to each of the boost operators:
\bealf{
\label{eq:lab frame wigner d sll}
\Delta n^{\textrm{th}}_{\ell m}(\nu)
&=\frac{\tau}{\gamma_{\rm p}} \sum_{\ell' m'}\sum_{\ell'' m''}\Bigg[\sum_{m_1}D_{m m_1}^{\ell}\Bb{-1}{}{m_1}{\ell\ell'}(\nu,-\beta_{\rm p})[D_{m' m_1}^{\ell'}]^*\Bigg]\,\Sll{\textrm{th}}{\ell'}(\nu,\The)
\nonumber\\
&\qquad\qquad\qquad\qquad\qquad \times
\Bigg[\sum_{m_2}D_{m' m_2}^{\ell'}\Bb{0}{}{m_2}{\ell'\ell''}(\nu,\beta_{\rm p})[D_{m'' m_2}^{\ell''}]^*\Bigg]\,\,n_{\ell'' m''}
(\nu).
}
Using the orthogonality of Wigner-D functions given by $\sum_{m'}[D_{m' m_1}^{\ell'}]^* D_{m' m_2}^{\ell'}=\delta_{m_1 m_2}$
we then obtain
\bealf{
\label{eq:lab frame wigner d sll simplified}
\Delta n^{\textrm{th}}_{\ell m}(\nu)
&
=\frac{\tau}{\gamma_{\rm p}} \sum_{\ell'}\sum_{\ell'' m''} \sum_{m_1}D_{m m_1}^{\ell}\Bb{-1}{}{m_1}{\ell\ell'}(\nu,-\beta_{\text{p}})\,\Sll{\textrm{th}}{\ell'}(\nu,\The)
\, \Bb{0}{}{m_1}{\ell'\ell''}(\nu,\beta_{\rm p})[D_{m'' m_1}^{\ell''}]^*\,\,n_{\ell'' m''}
(\nu)
\nonumber\\
&=\tau\sum_{\ell'}\sum_{\ell'' m''} \sum_{m_1}\left\{D_{m m_1}^{\ell}\Dbo{}{\ell \ell' \ell''}{m_1}(\nu,\beta_{\text{p}})\,[D_{m'' m_1}^{\ell''}]^*\right\}\Sll{\textrm{th}}{\ell'}(\nu,\The)\,\,n_{\ell'' m''}(\nu)
\nonumber\\
&=\tau\sum_{\ell'}\sum_{\ell'' m''}\, \Dbo{}{\ell \ell' \ell''}{mm''}(\nu,\boldsymbol{\beta}_{\text{p}})\,\Sll{\textrm{th}}{\ell'}(\nu,\The)\,n_{\ell'' m''}(\nu),
}
where in the second line we have replaced the boost operators with the corresponding Doppler operator (allowed since all operators commute) and in the third line we have implicitly defined the general direction Doppler operator $\Dbo{}{\ell \ell' \ell''}{mm''}(\nu,\boldsymbol{\beta}_{\text{p}})$ [again dropping the Doppler-weight in the notation]. This result could also be obtained directly from Eq.~\eqref{eq:sph harm coeff transform}, however, we have spelled out the intermediate steps to emphasise how the components come together. The remaining Wigner-D functions can be expressed through spin-weighted spherical harmonics as
\bealf{
\label{eq:wigner_d_spherical}
D_{m m'}^\ell(\phi,\theta,\psi)&\equiv
(-1)^{m'}\sqrt{\frac{4\pi}{2\ell+1}}\,{_{-m'}}Y^*_{\ell m}(\theta, \phi)\,\expf{-im'\psi},
}
meaning that with Eq.~\eqref{eq:lab frame wigner d sll simplified} the final relativistic SZ signal in the lab frame becomes
\bsub
\label{eq:lab frame final sll}
\bealf{
\Delta n_{\textrm{th}}(\nu, \vgh)
&
=\tau \sum_{\ell m}\sum_{\ell'' m''}Y_{\ell m}(\vgh)\,\Sll{m m''\,\textrm{th}}{\ell \ell''}(\nu,\The, \hat{\boldsymbol{\beta}}_{\rm p})\,n_{\ell''m''}(\nu),
\\
\Sll{m m''\,\textrm{th}}{\ell \ell''}(\nu,\The, \hat{\boldsymbol{\beta}}_{\rm p})&=\sum_{\ell'}\sum_{m_1} 
\frac{4 \pi\,{}_{-m_1}Y_{\ell m}^*(\hat{\boldsymbol{\beta}}_{\rm p})\Dbo{}{\ell \ell' \ell''}{m_1}(\nu,\beta_{\rm p})\Sll{\,\textrm{th}}{\ell'}(\nu,\The)\,{}_{-m_1}Y_{\ell'' m''}(\hat{\boldsymbol{\beta}}_{\rm p})}{\sqrt{(2\ell+1)(2\ell''+1)}}
}
\esub
where we have replaced the multipole coefficients with the full spectral distortion $\Delta n = \sum_{\ell m}Y_{\ell m}\Delta n_{\ell m}$. This final general form makes the angular dependence explicit through the spin-weighted spherical harmonics. Note that the dependence on the roll angle $\psi$ cancels as a result of rotating to be collinear with $\boldsymbol{\hat{z}}$, as rotation about this axis should not change the physics. At this stage, it is important to note that Eq.~\eqref{eq:lab frame final sll} describes the relativistic SZ effect (in the single-scattering limit) at all orders in $\The$, all orders in $\beta_{\text{p}}$ and even for a general unscattered (unpolarized) photon field.

\subsection{Monopole scattering}
\label{app: monopole scattering}
The incoming radiation field relevant to the SZ effects is the CMB, which is well-described by an isotropic Planckian distribution. Explicitly, 
\bealf{
\label{eq:isotropic pl}
n_0(\nu)=Y_{00}\,n_{00}(\nu)=\frac{1}{\expf{x}-1} \equiv n^{\textrm{Pl}}(\nu)
.
}
By inserting this distribution into Eq.~\eqref{eq:lab frame final sll}, the sums over $\ell''$ and $m''$ immediately collapse. Additionally, ${}_{-m_1}Y_{\ell'' m''}$ is zero for $|m_1|>\ell''$, enforcing $m_1=0$. This eliminates azimuthal mixing, giving 
\bsub
\label{eq:lab frame final sll_II}
\bealf{
\Delta n_{\textrm{th}}(\nu, \vgh)
&
=\tau \sum_{\ell m} Y_{\ell m}(\vgh)\,\Sll{m 0\,\textrm{th}}{\ell 0}(\nu,\The, \hat{\boldsymbol{\beta}}_{\rm p})\,n_{00}(\nu),
\\
\Sll{m 0\,\textrm{th}}{\ell 0}(\nu,\The, \hat{\boldsymbol{\beta}}_{\rm p})&=\sum_{\ell'}\frac{\sqrt{4 \pi}\,{}_{0}Y_{\ell m}^*(\hat{\boldsymbol{\beta}}_{\rm p})\Dbo{}{\ell \ell' 0}{0}(\nu,\beta_{\rm p})\,\Sll{\,\textrm{th}}{\ell'}(\nu,\The)}{\sqrt{2\ell+1}}.
}
\esub
As a consequence, we can use the addition theorem for spherical harmonics,
\bealf{
\label{eq: addition theorem}
\sum_{m=-\ell}^{\ell}Y_{\ell m}(\vgh)\,{}_{0}Y^*_{\ell m}(\vbh_{\text{p}})=\frac{2\ell+1}{4\pi}P_\ell(\vgh\cdot\vbh_{\text{p}}),
}
where $P_{\ell}$ denotes the Legendre polynomials of degree $\ell$. From this, we then obtain the expression
\bsub
\label{eq:sl combined}
\bealf{
\label{eq:sl combined I}
\Delta n_{\textrm{th}}(\nu, \vgh,\The, \boldsymbol{\beta}_{\rm p})
&
=\tau \, \Sll{}{\textrm{SZ}}(\nu,\vgh,\The, \boldsymbol{\beta}_{\rm p})\,n^{\textrm{Pl}}(\nu)
\\
\Sll{}{\textrm{SZ}}(\nu,\vgh,\The, \boldsymbol{\beta}_{\rm p})
&=
\sum_{\ell=0}^\infty \Sll{}{\ell}(\nu,\The, \beta_{\rm p}) P_\ell (\vgh\cdot\vbh_{\text{p}}),
\\
\label{eq:sl define}
\Sll{}{\ell}(\nu,\The, \beta_{\rm p})
&=
\sqrt{2\ell + 1} 
\sum_{\ell'} \Dbo{}{\ell \ell' 0}{0}(\nu, \beta_{\rm p})\,\Sll{\textrm{th}}{\ell'}(\nu,\The)
.
}
\esub
Here, we introduced the {\it relativistic SZ operator} $\Sll{}{\textrm{SZ}}(\nu,\vgh,\The, \boldsymbol{\beta}_{\rm p})$ and its Legendre decomposition required due to the intrinsic motion-induced anisotropy of the problem, with all thermal and kinematic effects encoded in the coefficients $\Sll{}{\ell}(\nu,\The,\beta_{\rm p})$. 

Equation~\eqref{eq:sl combined I} has the same form as Eq. (24) of \cite{Chluba2026SZ} (see also Appendix \ref{app : alt app}). It describes the relativistic SZ effect for a thermal distribution of electrons with bulk peculiar velocity $\vb_{\text{p}}$, still containing all orders in both $\The$ and $\beta_{\rm p}$ although it only includes scattering of the lab frame monopole spectrum (which need not be Planckian generally). Note that throughout this derivation we have considered the lab frame to be the same as the CMB rest frame, and we have not included effects that arise from the motion of the observer \citep{Chluba2005b}. 
Since we have obtained the spectral distortion in an equivalent form to that of \cite{Chluba2026SZ}, we are able to {\it directly} compare the SZ operators. Equating $\Sll{}{\ell}(\nu,\The, \beta_{\rm p})$ as given in Eq.~\eqref{eq:sl define} to the one of \citep{Chluba2026SZ} [see also Eq.~\eqref{eq:collect terms 2}], we have the identity
\bealf{
\label{eq:equate}
\int p^2 f_{\ell}(\gamma,\gamma_{\text{p}})\,\Sll{}{\ell}(\nu, p)\,\text{d}p\equiv \sqrt{2\ell + 1} \sum_{\ell'} \Dbo{}{\ell \ell' 0}{0}(\nu, \beta_{\rm p})\,\Sll{\textrm{th}}{\ell'}(\nu,\The),
}
where $\Sll{}{\ell}(\nu, p)$ is defined in Eq. (\ref{eq:collect terms 3}) and $f_{\ell}(\gamma,\gamma_{\text{p}})$ in Eq. (\ref{eq:collect terms 4}).\footnote{Using Eq.~\eqref{eq:Sll}, one can be tempted to write an equality for the integrand. Formally, however, this equality only holds for various order of the temperature, as we can show using {\tt Mathematica}. Therefore, only once the integrals over the distributions are taken does one obtain the identity but not for the integrands themselves.}
From this equation, we can see that by using successive boosts we have, in-effect, factorized the $\The$ and $\beta_{\text{p}}$ dependencies into two separate operators. Furthermore, this expression provides insight into the underlying symmetries of the Doppler operators which are further discussed in Section \ref{app: closed form scattering}.
We also validated this identity using {\tt Mathematica} to three orders in $\beta_{\text{p}}$ and four orders in $\The$. One important aspect is that in the new form, one can first compute all non-vanishing operators $\Dbo{}{\ell \ell' 0}{0}(\nu, \beta_{\rm p})$ at the considered order in $\beta_{\text{p}}$. This then determines the maximal value for $\ell'$ that has to be considered. Indeed $\ell'_{\rm max}$ is identical to the maximal order in $\beta_{\text{p}}$, e.g., $\ell'_{\rm max}=3$ for terms up to third order in $\beta_{\text{p}}$. The temperature order of the operators $\Sll{\textrm{th}}{\ell'}(\nu,\The)$ can then be chosen as required. Also, one can generate all terms $\Sll{}{\ell}(\nu,\The, \beta_{\rm p})$ as simple linear combinations of all these $\Sll{\textrm{th}}{\ell'}(\nu,\The)$. In contrast, working with the expressions given previously one has to carefully make sure that no term is missing especially because the expansions of the distribution functions contain terms $\beta_{\rm p}/\The$.
In Appendix~\ref{operator expansions app} we give all operators required to obtain the results up to $\mathcal{O}(\The^4)$ and $\mathcal{O}(\beta_{\rm p}^3)$. At the end we also give a link to where we provide a {\tt Mathematica} notebook to generate higher order terms.

\subsection{Kinematic corrections to $\mathcal{O}(\beta_{\rm p}^3)$}
\label{expansions}
We can now expand the SZ operator to obtain expressions for the kinematic correction to the SZ effect. Here, we will first calculate the kinematic corrections to third order in $\beta_{\rm p}$. Using {\tt Mathematica}, we can expand each of the scattering operators (at zeroth order in $\The$) to examine their operator structure. For $\ell=0,1,2$, we have, 
\bsub
\label{szoperatorsexpanded3rdorder}
\bealf{
\label{eq:S0 expansion}
\Sll{}{0}(\nu,p_{\text{p}})
&\approx
\frac{1}{3}\oDnu p_{\text{p}}^2 + \frac{7}{150}\oDnu(\oDnu-4)p_{\text{p}}^4
+\frac{11}{3150}\oDnu(\oDnu-4)(\oDnu-10)p_{\text{p}}^6,\\
\label{eq:S1 expansion}
\Sll{}{1}(\nu,p_{\text{p}})
&\approx
\oOnu p_{\text{p}} + \frac{1}{50}\oOnu\left(9 - 10\oOnu + 14\oDnu\right)p_{\text{p}}^3 
+
\frac{1}{1400} \oOnu\Big(-319+252\oOnu-280\oDnu
\nonumber\\
&\qquad\qquad
-28\oOnu\oDnu+44\oDnu^2\Big)p_{\text{p}}^5,
\\
\label{eq:S2 expansion}
\Sll{}{2}(\nu,p_{\text{p}})
&\approx
\frac{1}{30}\oOnu \left(24\oOnu+11\oDnu\right)p_{\text{p}}^2
+\frac{1}{210}\oOnu\left(48-64\oOnu-9\oDnu+19\oOnu\oDnu\right)p_{\text{p}}^4 
\nonumber\\
&\qquad\qquad
+\frac{1}{13230}\left(\oDnu-10\right)\Big(-888\oOnu-376\oDnu
+306\oOnu\oDnu+115\oDnu^2\Big)p_\text{p}^6
}
\esub
where $\oOnu$ is the aforementioned energy shift generator and $\oDnu=\oOnu^2-3\oOnu=x^{-2}\partial_xx^4\partial_x$ is the Kompaneets diffusion operator. $\Sll{}{\ell}$ is constructed from $\Dbo{}{\ell \ell' 0}{0}(\nu, \beta_{\rm p})$ and $\Sll{\textrm{th}}{\ell'}(\nu,\The)$, and the terms required to give the above expressions are given in Appendix \ref{operator expansions app}. The expressions for $\Sll{}{1}$ and $\Sll{}{2}$ are also given in Eq.~(28) of \cite{Chluba2026SZ}, and we find these to be in agreement once inserting $\oOnu$ and $\oDnu$ in their derivative forms. In regard to the structure of these scattering operators, we see that for $\ell=0$, we only have contributions from the diffusion operator, owing to the fact we are not mixing lab frame multipoles. However, for $\ell>0$ we have energy shift terms $\propto \oOnu$ appearing. As we see in Appendix~\ref{operator expansions app}, the thermal SZ operators $\Sll{\textrm{th}}{0}$ are described by $\oDnu$ only, hence the appearance of the energy shift generator is entirely due to the bulk motion of the electron cloud, as might be expected. 

Expanding the full SZ operator to $\ell = 3$, zeroth order in $\The$, we obtain the kinematic correction to the SZ effect to $\mathcal{O}(\beta_{\rm p}^3)$,
\bealf{
\label{eq:Ssz expansion}
\Sll{}{\textrm{kin}}(\nu,\The,\vgh,\boldsymbol{\beta}_\text{p})
&\approx 
\beta_\text{p} \mu_\text{p} \oOnu +
\frac{\beta_\text{p}^2}{3}\left[\oDnu+P_2(\mu_\text{p})\left(\frac{12}{5}\oOnu + \frac{11}{10}\oDnu\right)\right]
\\
\nonumber
&\qquad\qquad
+
\frac{\beta_\text{p}^3}{25}\Bigg[\mu_\text{p}\left(2\oOnu-5\oDnu+7\oOnu\oDnu\right)
+\frac{P_3(\mu_\text{p})}{6}\left(128\oOnu+45\oDnu+13\oOnu\oDnu\right)\Bigg]
,
}
which, again, is in agreement with \cite{Chluba2026SZ}. The first term is the classical kSZ effect \citep{Sunyaev1980}; terms $\propto \beta_\text{p}^2$ are leading order corrections due to monopole and quadrupole scattering. To our knowledge, $\beta_{\rm p}^3$ corrections have only previously been derived in \cite{Chluba2026SZ}, since these terms prove considerably more difficult to derive using conventional methods.\footnote{Note that these terms are usually negligible for typical cluster velocities.} 
\subsection{First order temperature correction}
\label{temp cor}
Next, we explicitly give the first order temperature correction to the kSZ.  Considering the operator expansions (see Appendix \ref{operator expansions app}), each thermally-averaged scattering operator, $\Sll{\textrm{th}}{\ell}(\nu,\The)$, up to $\ell=3$ contains corrections linear in $\The$. We also retain contributions of $\mathcal{O}(p_\text{p})$ which correspond to contributions at $\mathcal{O}(\beta_\text{p})$, i.e.,  $p_\text{p}\approx\beta_{\rm p}+\frac{\beta_{\rm p}^3}{2}+\mathcal{O}(\beta_\text{p}^5)$. The only elements containing contributions linear in $p_\text{p}$ are $\Dbo{}{100}{}(\nu, p_\text{p})$ and $\Dbo{}{110}{}(\nu, p_\text{p})$.
Hence, at first order in $\beta_\text{p}$, the first order temperature correction is
\bealf{
\label{eq:first order T}
\Sll{\The}{\textrm{kin}}&\approx\sqrt{3}\bigg[\Dbo{}{100}{}(\nu,\beta_{\rm p})\Sll{\textrm{th}}{0}(\nu,\The)+\Dbo{}{110}{}(\nu,\beta_{\rm p})\Sll{\textrm{th}}{1}(\nu,\The)\bigg]\,\mu_{\rm p}
\nonumber
\\
&
\approx \sqrt{3}\bigg[\frac{\oDnu}{\sqrt{3}}(\oOnu-1)\,\The\,\beta_{\rm p}+\frac{2\oOnu}{5\sqrt{3}}(1+\oDnu)\,\The\,\beta_{\rm p}\bigg]\,\mu_{\rm p}
\nonumber
\\
&
\approx \bigg[\oDnu(7\oOnu-1)+2\oOnu\bigg]\frac{\The\,\beta_{\rm p}\,\mu_{\rm p}}{5},
}
which reproduces Eq.~(29) of \cite{Chluba2026SZ}, as well as Eq.~(28) of \citep{Chluba2012moments} after replacing $\tau$ with the cloud rest frame optical depth $\tau=\tau'/(1-\beta_\text{p}\mu_\text{p})$ \citep[see section 4.4 of ][for details]{Chluba2012SZpack}. Further cross checks confirm this expression agrees with $C_1$, as given in Eq.~(29) of \cite{Nozawa1998SZ}. The same method can, of course, be employed to derive the second order temperature corrections and beyond simply by considering which operator elements contribute to the signal at that order. As stressed above, one can obtain all expressions as linear combination of $\Sll{\textrm{th}}{\ell'}(\nu,\The)$, which greatly simplifies the overall computation.
This demonstrates the systematic nature of the boost operator approach, where the dependencies on $\The$ and $\beta_{\rm p}$ are treated separately and can therefore be expanded independently to any order.
\section{Closed form expression for the thermal SZ scattering operator}
\label{app: closed form scattering}
As we have seen, the SZ operators we obtained must be expanded in orders of $\The$ and $\beta_\text{p}$ to compute the SZ signal, thus it would be advantageous to obtain closed form expressions for these expansions. Another motivation for searching for closed forms is attempting to better understand the structure of the Doppler operators which, thanks to the boost operator approach, simply and completely describe the effect of scattering processes on the photon distribution. The full relativistic SZ operator, given in Eq.~\eqref{eq:sl combined}, has a complex operator structure, so instead we focus on the simpler case of the thermal SZ scattering operator (for the monopole only), which is defined by 
\bealf{
\label{eq:thermal_scattering_op}
\Sll{\textrm{th}}{0}(\nu', \The) &= \Dbo{-1}{0 0 0}{}(\nu', \beta) + \frac{1}{10}\Dbo{-1}{0 2 0}{}(\nu', \beta) - 1.
}
In this section we will provide a closed form expansion of ${}^{-1}\hat{\mathcal{D}}_{000}$. Following this we obtain a closed-form for ${}^{-1}\hat{\mathcal{D}}_{020}$ through inspection of coefficients, and hence a closed form for the thermal (monopole) SZ operator. Additional closed form expansions, which are not relevant to the thermal scattering operator are given in Appendix \ref{app:additional explicit}, but may give hints about the more general properties.

\subsection{Closed form for ${}^d\hat{\mathcal{D}}_{000}$}
\label{closed form derivation}
\noindent
We begin by considering the first term in Eq.~\eqref{eq:thermal_scattering_op}, but for a general Doppler weight $d$, i.e. $\Dbo{d}{0 0 0}{}$\footnote{Note that we are considering the $m$-averaged Doppler operator, although there is no explicit difference for the Doppler operator and its $m$ averaged counterpart when at least one $\ell$ index is zero.}. Expressed in terms of boost operators,
\bealf{
\label{eq:D000}
{}^d\hat{\mathcal{D}}_{000}&=\frac{1}{\gamma}{}^d\hat{\mathcal{B}}_{00}^{0}(-\beta)\,{}^0\hat{\mathcal{B}}_{00}^{0}(\beta)
=\frac{1}{\gamma}{}^{d+\oOnu}\mathcal{K}_{00}^{0}(\beta)\,{}^{2-\oOnu}\mathcal{K}_{00}^{0}(\beta),
}
where, in the second line, we have used ${}^{d}\hat{\mathcal{B}}^{m}_{\ell \ell'}(\nu,\beta)
\equiv {}^{d+\oOnu}\mathcal{K}^{m}_{\ell \ell'}(-\beta)$ and ${}^{d}_{s}\mathcal{K}_{\ell \ell'}^{m}\,(\beta)={}^{2-d}_{s}\mathcal{K}_{\ell' \ell}^{m}\,(-\beta)$ \citep{ChlubaBO25}. 
We continue by inserting the explicit form for the aberration kernel, as given in Eq.~\eqref{eq: Kernel example}, into the above equation. Defining $p_+=(\gamma+p)$ and $p_-=(\gamma-p)$, we have
\bealf{
\label{eq:D000_expanded}
{}^d\hat{\mathcal{D}}_{000}&=\frac{1}{\gamma} \Bigg[\frac{p_+^{1-\oOnu-d}-p_-^{1-\oOnu-d}}{2p\,(1-\oOnu-d)}\Bigg]\Bigg[\frac{p_+^{\oOnu-1}-p_-^{\oOnu-1}}{2p\,(\oOnu-1)}\Bigg]\nonumber\\
&
=\frac{1}{4\gamma p^2\,(1-\oOnu-d)(\oOnu-1)}\Bigg(p_+^{-d}+p_-^{-d}-p_+^{2-2\oOnu-d}-p_-^{2-2\oOnu-d}\Bigg)\nonumber\\
&
=\frac{1}{4\gamma p^2\,(1-\oOnu-d)(\oOnu-1)}\Bigg(\expf{-d\ln{p_+}}+\expf{d\ln{p_+}}
-\expf{(2-2\oOnu-d)\ln{p_+}}-\expf{-(2-2\oOnu-d)\ln{p_+}}\Bigg)\nonumber\\
&
=\frac{1}{2\gamma p^2\,(1-\oOnu-d)(1-\oOnu)}\Bigg\{\cosh{\Big[(2-2\oOnu-d)\ln{p_+}\Big]}
-\cosh{\Big[d\ln{p_+}\Big]}\Bigg\},
}
Where we have used the fact that $p_+p_-=1$. We then recognise that $\ln{p_+}=\ln{(p+\sqrt{1+p^2})}=\textrm{arsinh}(p)$, which allows us to write
\bealf{
\label{eq:D000_expanded2}
{}^d\hat{\mathcal{D}}_{000}
&
=\frac{1}{2\gamma p^2\,(1-\oOnu-d)(1-\oOnu)}\Bigg\{\cosh{\Big[(2-2\oOnu-d)\,\textrm{arsinh}(p)\Big]}
-\cosh{\Big[d\,\textrm{arsinh}(p)\Big]}\Bigg\}.
}
A detailed derivation of a closed form expansion for the function $\cosh{[k\,\textrm{arsinh}(p)]/\sqrt{1+p^2}}$ is given in Appendix \ref{app: closed form arsinh derivation}. Inserting this closed form into our expression for ${}^d\hat{\mathcal{D}}_{000}$ above,
we have
\bealf{
\label{eq:D000_expanded3}
{}^d\hat{\mathcal{D}}_{000}
&=\frac{1}{2(1-\oOnu-d)(1-\oOnu)}\sum_{n=0}^{\infty}\frac{p^{2n-2}}{(2n)!}\Bigg\{\prod_{m=0}^{n-1}\Big[(2-2\oOnu-d)^2
-(2m+1)^2\Big]-\prod_{m=0}^{n-1}\left[d^2-(2m+1)^2\right]\Bigg\}
\nonumber\\
&=
\frac{1}{\oDnu+2+(1+d)\,(\oOnu-1)}\sum_{n=0}^{\infty}\frac{2^{2n+1}\,p^{2n}}{[2(n+1)]!}\Bigg[\prod_{m=0}^{n}\Bigg\{\oDnu+2+(1+d)\,(\oOnu-1)
\\ \nonumber
&\qquad\qquad\qquad
+\frac{d^2-1}{4}-m(m+1) \Bigg\}
-\prod_{m=0}^{n}\left\{\frac{d^2-1}{4}-m(m+1)\right\}\Bigg],
}
where in the second line we have reordered the sums and collected the powers of $p$, expanding the quadratic in $m$. We note that the leading factor $[\oDnu+2+(1+d)\,(\oOnu-1)]^{-1}$ always drops out in these expressions. This can be seen when defining $a=\oDnu+2+(1+d)\,(\oOnu-1)$ and $a_m=\frac{d^2-1}{4}-m(m+1)$. With this, we can write the products as
\bealf{
\prod_{m=0}^{n}\left(a+a_m\right)-\prod_{m=0}^{n} a_m=a\sum_{j=0}^n \left\{\prod_{m=0}^{j-1}\left(a+a_m\right)\prod_{m=j+1}^{n} a_m \right\},
}
where $\prod_{m=0}^{-1}[\ldots]=1$. However, this alternative version of writing things is not obviously simpler for applications.
From the above expression, it is immediately clear that for $d=-1$ we then have
\bealf{
\label{eq:D000_expanded2_dm1}
{}^{-1}\hat{\mathcal{D}}_{000}
&=
\frac{1}{\oDnu+2}\sum_{n=0}^{\infty}\frac{2^{2n+1}\,p^{2n}}{[2(n+1)]!}\prod_{m=0}^{n}\left\{\oDnu+2-m(m+1)\right\}
\nonumber\\
&
=
\sum_{n=0}^{\infty}\frac{2^{2n+1}\,p^{2n}}{[2(n+1)]!}\,\prod_{m=0}^{n-1
}\left\{\oDnu-m(m+3)\right\},
}
and so we have obtained a closed form expansion for the desired ${}^{-1}\hat{\mathcal{D}}_{000}$, together with a more general closed form for ${}^d\hat{\mathcal{D}}_{000}$. Written in this way we can appreciate the operator structure. Specifically that dependence on $\oOnu$ drops out for $d=-1$. This might be expected, given the knowledge that the diffusion operator for $d=-1$ describes Thomson scattering with no recoil effects, and because we are coupling the same initial and final multipoles\footnote{This is a poor example of this fact, since we are going from $\ell''=0$ to $\ell=0$ through the intermediate multipole $\ell'=0$, however we will find that the $\oOnu$ dependence will always drop out when $\ell''=\ell$ (for $d=-1$).}. Looking back at Eq.~\eqref{eq:D000_expanded3}, we see that the further $d$ deviates from $-1$, the greater the contribution of $\oOnu$. This can be understood as the increased energy exchange for quantities which transform with higher Doppler weight (e.g. higher orders in recoil).
\subsection{Monopole scattering SZ operator}
\label{monopolszresult}
\noindent
Now that we have obtained an expression for ${}^{-1}\hat{\mathcal{D}}_{000}$, we turn our attention to the second term in Eq.~\eqref{eq:thermal_scattering_op}: ${}^{-1}\hat{\mathcal{D}}_{020}$. Comparing the expansion of ${}^{-1}\hat{\mathcal{D}}_{020}$ with ${}^{-1}\hat{\mathcal{D}}_{000}$ in {\tt Mathematica}, we observe an identical pattern in the structure of the operators, only with different numerical coefficients. Inspection of these coefficients yields
\bealf{
\label{eq:D020_expanded2_dm1}
{}^{-1}\hat{\mathcal{D}}_{020}
&=
\sum_{n=0}^{\infty}\frac{2^{2n+1}\,p^{2n}}{[2(n+1)]!}\,\frac{5 n(n-1)}{(n+3)(n+2)}\,\prod_{m=0}^{n-1}\left\{\oDnu-m(m+3)\right\}.
}
It is important to note that we do not have a formal proof for the above expression, however we have confirmed that it is correct to $\mathcal{O}(p^{20})$. We suspect it can again be proven by using the explicit forms for the required aberration kernels and boost operators together with the relevant recurrence relations, however, we leave a deeper investigation for later.

Combining Eq.~\eqref{eq:D000_expanded2_dm1} with Eq.~\eqref{eq:D020_expanded2_dm1}, we have a closed-form expression for tSZ scattering operator for monopole scattering,
\bealf{
\label{eq:Sth0_expanded}
\Sll{\textrm{th}}{0}(\nu', \beta) 
&= 
\sum_{n=0}^{\infty}\frac{2^{2n+1}\,p^{2n}}{[2(n+1)]!}\left[1+\frac{n(n-1)}{2(n+3)(n+2)}\right]\,\prod_{m=0}^{n-1}\left\{\oDnu-m(m+3)\right\}-1
\nonumber\\
&=
\sum_{n=1}^{\infty}\frac{2^{2n}\,p^{2n}}{[2(n+1)]!}\,\frac{3[(n+3)n+4]}{(n+3)(n+2)}\,\prod_{m=0}^{n-1}\left\{\oDnu-m(m+3)\right\},
}
where we have combined the terms into a single coefficient for $p^{2n}$ and in the second line, the $n=0$ term of the sum has cancelled the $-1$. To compute the thermal average, we use the moments \citep[see][]{CSpack2019} 
\bealf{
\label{app:MMrMB_k}
\left<p^{2n}\right>&=\frac{2\,(2\The)^{n} K_{n+2}(1/\The)}{\sqrt{\pi}K_2(1/\The)}\,\Gamma\left(n+\frac{3}{2}\right)
=
\frac{(2\The)^{n} K_{n+2}(1/\The)}{K_2(1/\The)}\,\frac{[2(n+1)]!}{2^{2n+1}(n+1)!},
}
which implies
\bealf{
\label{eq:S0_final}
\Sll{\textrm{th}}{0}(\nu', \The) 
&=
\sum_{n=1}^{\infty}\,\frac{2^{2n}}{[2(n+1)]!}\,\frac{(2\The)^{n} K_{n+2}(1/\The)}{K_2(1/\The)}\,\frac{[2(n+1)]!}{2^{2n+1}(n+1)!}
\nonumber\\
&\qquad \times
\frac{3[(n+3)n+4]}{(n+3)(n+2)}\,\prod_{m=0}^{n-1}\left\{\oDnu-m(m+3)\right\}
\nonumber\\
&=
\sum_{n=1}^{\infty}\frac{\The^{n} K_{n+2}(1/\The)}{K_2(1/\The)}\frac{2^{n}\,3[(n+3)n+4]}{2(n+3)!}\,\prod_{m=0}^{n-1}\left\{\oDnu-m(m+3)\right\},
}
as the final closed-form expansion. The modified Bessel functions of second kind that arise from the thermal average are themselves functions of $\The$ and must also be expanded (although this is simply taken care of in {\tt Mathematica}). Simple recurrence relations to generate all the coefficients in orders of $\The$ can be found by inspecting the differential equation for the modified Bessel function, however, we omit the solutions here given that {\tt Mathematica} can generate these terms easily.

We note, however, that even with this all orders in $\The$ expression one cannot overcome the convergence issues for the SZ effect inherent to the asymptotic nature of the expansion \citep{Itoh98,Chluba2012SZpack}. This derives from the fact that the Wien tail of the Planckian goes as $n^{\textrm{Pl}}(x)\simeq \expf{-x}$ for $x \gg 1$. Since the frequency derivatives of this Planckian are steep, a small shift in frequency causes a large fractional change in the spectrum. That is, the characteristic frequency interval over which they change rapidly is smaller than the frequency change resulting from the scattering process. The action of repeated differential operators in higher-order terms amplifies the large derivatives causing oscillatory overcorrections and poor convergence at high frequencies. Thus, the expression becomes non-perturbative even if one directly uses the exact moment expressions as they are (i.e., without performing any expansion in $\The$).

\subsection{Going beyond monopole scattering}
\label{beyond mono}
If we wish to obtain closed-form expressions for the relativistic SZ operator beyond the monopole, we must obtain closed forms for $\Dbo{-1}{\ell0\ell}{}$ and $\Dbo{-1}{\ell2\ell}{}$, with $\ell\neq0$. As we did for Eq.~\eqref{eq:D020_expanded2_dm1},
when inspecting operators like $\Dbo{-1}{101}{}$ and $\Dbo{-1}{202}{}$ in {\tt Mathematica}, we find a consistent operator structure, with differing numerical coefficients. By generalising this pattern, we obtain
\bealf{
\label{eq:Dl0l_expanded2_dm1}
{}^{-1}\hat{\mathcal{D}}_{\ell0\ell}
&=
\frac{(-1)^{\ell}(2\ell+1)}{\oDnu+2-\ell(\ell+1)}
\sum_{n=0}^{\infty}\frac{2^{2n+1}\,p^{2n}}{[2(n+1)]!}\,\frac{(n)_\ell}{(n+2)^{(\ell)}}
\prod_{m=0}^{n}\left\{\oDnu+2-m(m+1)\right\},
}
where $(n)_\ell=n(n-1)(n-2)\cdots(n-\ell+1)$ and $n^{(\ell)}=n(n+1)(n+2)\cdots(n+\ell-1)$ denote the falling and rising factorials (or Pochammer function), respectively. Again, we have no formal proof for this expression, however we have confirmed up to $\ell=8$ and to $\mathcal{O}(p^{20})$, through {\tt Mathematica}, that it reproduces the operator expansions identically.

On the other hand, we were unable to find any discernible pattern when inspecting terms such as $\Dbo{-1}{121}{}$ and $\Dbo{-1}{222}{}$. This deviation from what we have previously observed is due to the operators being $m$-averaged. Until now, the operators we have considered have had at least one $\ell$ index equal to zero, resulting in only one contribution ($m=0$) to the $m$-averaged operator. For Doppler operators with $\ell>0$ for each of its indices, only the contribution with maximal $m$, i.e. $|m|=\ell_{\text{min}}$, observes any resemblance of a tidy pattern; all other contributions introduce a complicated operator structure. When we focus on the maximal $m$ operators in {\tt Mathematica}, for example $\Dbo{-1}{121}{\pm1}$, we find
\bealf{
\label{eq:D121m1_expanded2_dm1}
{}^{-1}\hat{\mathcal{D}}_{121}^{\pm1}
&=
-\sum_{n=1}^{\infty}\frac{1}{5}\prod_{k=1}^{n-1}\frac{2(2+k)}{k(5+k)(5+2k)}\prod_{m=0}^{n-1}[\oDnu-4-m(m+5)]p^{2n},
}
However, to obtain an expression for the full $m$-averaged operators, further research is needed on the properties of the boost operator. We leave this exploration to future work.

\section{Conclusions}
\label{sec:Conc}
In this work, we have used the boost operator approach \citep{ChlubaBO25} to present a novel derivation of the relativistic Sunyaev-Zeldovich effect including both kinematic and temperature corrections (see Section \ref{kinematic sz main derivation}), with the SZ signal described exactly by the SZ operator in Eq.~\eqref{eq:sl combined}, at all orders in the electron temperature $\The$ and the peculiar velocity $\beta_\text{p}$. This provides a simplification from previous work, including that of \citep{Chluba2026SZ}, by reordering the required operations so that the thermal averaging is performed in the electron cloud rest frame \textit{before} incorporating the peculiar motion of the electron cloud. The difference that results from this reordering is made explicit in Eq.~\eqref{eq:equate}, which highlights the operator factorization. We confirmed this identity using {\tt Mathematica}.
By expanding the SZ operator in {\tt Mathematica}, we obtain expressions for the scattering operators for multipoles $\ell=0,1,2$ [see Eq.~\eqref{szoperatorsexpanded3rdorder}], the kinematic correction at third order in $\beta_\text{p}$ [Eq.~\eqref{eq:Ssz expansion}], and the first order temperature correction [see Eq.~\eqref{eq:first order T}], all of which are confirmed to be in agreement with the literature, evidencing the accuracy of this approach. Higher order corrections can be easily generated using our expressions. In this derivation, we treat the lab frame as identical to the CMB rest frame and do not consider corrections due to the motion of the observer \citep{Chluba2005b}, which could also be considered using the boost operator approach. Similarly, the approach can be readily extended to simplify the treatment of kinematic corrections to the polarized SZ effect \citep{Rosenberg2025prSZ}.

We also derived a closed-form expression for the relativistic (monopole scattering) thermal SZ operator (see Section \ref{app: closed form scattering}) in orders of $\The$ [see Eq.~\eqref{eq:S0_final}]. This was achieved through first deriving a closed-form expansion for ${}^{-1}\hat{\mathcal{D}}_{000}$ [see Eq.~\eqref{eq:D000_expanded2_dm1}], and then inspecting the operator structure to obtain ${}^{-1}\hat{\mathcal{D}}_{020}$ [see Eq.~\eqref{eq:D020_expanded2_dm1}]. However we could not succeed in finding a formal proof for the latter. This closed form can be applied in {\tt Mathematica} to efficiently compute the tSZ operator at any order in $\The$. However, even when going to extreme orders in $\The$, the expansion becomes non-convergent at high frequency, a well-known feature that is also discussed in \citep{Chluba2012SZpack}. 

The search for a closed-form expressions for the operators with $\ell>0$ proved to be more challenging (see Section \ref{beyond mono}). The underlying structure of the Doppler operator elements suggests a closed form expression for the kSZ scattering operator may exist. Although closed form representations for some of the lowest multipole elements have already been obtained, an incomplete understanding of the full symmetries of the Doppler operators prevents more general expressions from being found. We suspect that by using the properties of the boost operator one can unravel a deeper structure for the Doppler operators, an exciting exploration that is left to future work.


\vspace{2mm}

\noindent
{\it Data Availability Statement}: {\tt Mathematica} files to reproduce some of the key results are available at \url{www.chluba.de/Mathematica}.

\vspace{-2mm}
\section*{Acknowledgments}
This work was supported by the UKSA grant: LiteBIRD UK ST/Y005945/1. The authors would like to thank Boris Bolliet and Adrien Cristian for helpful comments on the manuscript and pointing out typos in some of our expressions.

{
\bibliographystyle{plain}
\bibliography{Lit-2025-edit}

@ARTICLE{Remazeilles2019CellrSZ,
       author = {{Remazeilles}, Mathieu and {Bolliet}, Boris and {Rotti}, Aditya and {Chluba}, Jens},
        title = "{Can we neglect relativistic temperature corrections in the Planck thermal SZ analysis?}",
      journal = {\mnras},
         year = 2019,
        month = mar,
       volume = {483},
       number = {3},
        pages = {3459-3464},
          doi = {10.1093/mnras/sty3352},
archivePrefix = {arXiv},
       eprint = {1809.09666},
 primaryClass = {astro-ph.CO},
       adsurl = {https://ui.adsabs.harvard.edu/abs/2019MNRAS.483.3459R}
}

@ARTICLE{Lee2022SZ,
       author = {{Lee}, Elizabeth and {Anbajagane}, Dhayaa and {Singh}, Priyanka and {Chluba}, Jens and {Nagai}, Daisuke and {Kay}, Scott T. and {Cui}, Weiguang and {Dolag}, Klaus and {Yepes}, Gustavo},
        title = "{A multisimulation study of relativistic SZ temperature scalings in galaxy clusters and groups}",
      journal = {\mnras},
         year = 2022,
        month = dec,
       volume = {517},
       number = {4},
        pages = {5303-5324},
          doi = {10.1093/mnras/stac2781},
archivePrefix = {arXiv},
       eprint = {2207.05834},
 primaryClass = {astro-ph.CO},
       adsurl = {https://ui.adsabs.harvard.edu/abs/2022MNRAS.517.5303L}
}

@ARTICLE{Lee2020SZ,
       author = {{Lee}, Elizabeth and {Chluba}, Jens and {Kay}, Scott T. and {Barnes}, David J.},
        title = "{Relativistic SZ temperature scaling relations of groups and clusters derived from the BAHAMAS and MACSIS simulations}",
      journal = {\mnras},
         year = 2020,
        month = apr,
       volume = {493},
       number = {3},
        pages = {3274-3292},
          doi = {10.1093/mnras/staa450},
archivePrefix = {arXiv},
       eprint = {1912.07924},
 primaryClass = {astro-ph.CO},
       adsurl = {https://ui.adsabs.harvard.edu/abs/2020MNRAS.493.3274L}
}

@ARTICLE{DiMascolo2025AtLAST,
       author = {{Di Mascolo}, Luca and {Perrott}, Yvette and
                 {Mroczkowski}, Tony and {Raghunathan}, Srinivasan and
                 {Andreon}, Stefano and {Ettori}, Stefano and
                 {Simionescu}, Aurora and {van Marrewijk}, Joshiwa and
                 {Cicone}, Claudia and {Lee}, Minju and {Nelson}, Dylan and
                 {Sommovigo}, Laura and {Booth}, Mark and {Klaassen}, Pamela and
                 {Andreani}, Paola and {Cordiner}, Martin A. and
                 {Johnstone}, Doug and {van Kampen}, Eelco and {Liu}, Daizhong and
                 {Maccarone}, Thomas J. and {Morris}, Thomas W. and
                 {Orlowski-Scherer}, John and {Saintonge}, Am{\'e}lie and
                 {Smith}, Matthew and {Thelen}, Alexander E. and
                 {Wedemeyer}, Sven},
        title = "{Atacama Large Aperture Submillimeter Telescope (AtLAST)
                  science: Resolving the hot and ionized Universe through
                  the Sunyaev-Zeldovich effect}",
      journal = {Open Research Europe},
         year = 2025,
       volume = {4},
          eid = {113},
        pages = {113},
          doi = {10.12688/openreseurope.17449.2},
archivePrefix = {arXiv},
       eprint = {2403.00909},
 primaryClass = {astro-ph.CO},
       adsurl = {https://ui.adsabs.harvard.edu/abs/2025ORE.....4..113D}
}

@ARTICLE{Coulton2026rSZ,
       author = {{Coulton}, William R. and {Duivenvoorden}, Adriaan J. and {Atkins}, Zachary and {Battaglia}, Nicholas and {Battistelli}, Elia Stefano and {Bond}, J. Richard and {Cai}, Hongbo and {Calabrese}, Erminia and {Choi}, Steve K. and {Crowley}, Kevin T. and {Devlin}, Mark J. and {Dunkley}, Jo and {Ferraro}, Simone and {Guan}, Yilun and {Herv{\'\i}as-Caimapo}, Carlos and {Hill}, J. Colin and {Hilton}, Matt and {Hincks}, Adam D. and {Kosowsky}, Arthur and {Madhavacheril}, Mathew S. and {van Marrewijk}, Joshiwa and {McCarthy}, Fiona and {Moodley}, Kavilan and {Mroczkowski}, Tony and {Niemack}, Michael D. and {Page}, Lyman A. and {Partridge}, Bruce and {Schaan}, Emmanuel and {Sehgal}, Neelima and {Sherwin}, Blake D. and {Sif{\'o}n}, Crist{\'o}bal and {Spergel}, David N. and {Staggs}, Suzanne T. and {Van Engelen}, Alexander and {Vavagiakis}, Eve M. and {Wollack}, Edward J.},
        title = "{Atacama Cosmology Telescope: A measurement of galaxy cluster temperatures through relativistic corrections to the thermal Sunyaev-Zeldovich effect}",
      journal = {\prd},
         year = 2026,
        month = feb,
       volume = {113},
       number = {4},
          eid = {043520},
        pages = {043520},
          doi = {10.1103/n7p5-pc66},
archivePrefix = {arXiv},
       eprint = {2410.19046},
 primaryClass = {astro-ph.CO},
       adsurl = {https://ui.adsabs.harvard.edu/abs/2026PhRvD.113d3520C}
}

@ARTICLE{Remazeilles2025MNRAS,
       author = {{Remazeilles}, Mathieu and {Chluba}, Jens},
        title = "{Evidence for relativistic Sunyaev-Zeldovich effect in Planck CMB maps with an average electron-gas temperature of $T_{\rm e}\simeq$ 5 keV}",
      journal = {\mnras},
         year = 2025,
        month = apr,
       volume = {538},
       number = {3},
        pages = {1576-1586},
          doi = {10.1093/mnras/staf384},
archivePrefix = {arXiv},
       eprint = {2410.02488},
 primaryClass = {astro-ph.CO},
       adsurl = {https://ui.adsabs.harvard.edu/abs/2025MNRAS.538.1576R}
}

@ARTICLE{Butler2022rSZ,
       author = {{Butler}, Victoria L. and {Feder}, Richard M. and
                 {Daylan}, Tansu and {Mantz}, Adam B. and {Mercado}, Dale and
                 {Monta{\~n}a}, Alfredo and {Portillo}, Stephen K. N. and
                 {Sayers}, Jack and {Vaughan}, Benjamin J. and
                 {Zemcov}, Michael and {Zitrin}, Adi},
        title = "{Measurement of the Relativistic Sunyaev-Zeldovich
                  Correction in RX J1347.5-1145}",
      journal = {\apj},
         year = 2022,
        month = jun,
       volume = {932},
       number = {1},
          eid = {55},
        pages = {55},
          doi = {10.3847/1538-4357/ac6c04},
archivePrefix = {arXiv},
       eprint = {2110.13932},
 primaryClass = {astro-ph.CO},
       adsurl = {https://ui.adsabs.harvard.edu/abs/2022ApJ...932...55B}
}

@ARTICLE{Ruan2013mergerSZ,
       author = {{Ruan}, John J. and {Quinn}, Thomas R. and {Babul}, Arif},
        title = "{The observable thermal and kinetic Sunyaev-Zel'dovich
                  effect in merging galaxy clusters}",
      journal = {\mnras},
         year = 2013,
        month = jul,
       volume = {432},
       number = {4},
        pages = {3508-3519},
          doi = {10.1093/mnras/stt701},
archivePrefix = {arXiv},
       eprint = {1304.6088},
 primaryClass = {astro-ph.CO},
       adsurl = {https://ui.adsabs.harvard.edu/abs/2013MNRAS.432.3508R}
}

@ARTICLE{CSpack2019,
       author = {{Sarkar}, Abir and {Chluba}, Jens and {Lee}, Elizabeth},
        title = "{Dissecting the Compton scattering kernel I: Isotropic media}",
      journal = {\mnras},
         year = 2019,
        month = dec,
       volume = {490},
       number = {3},
        pages = {3705-3726},
          doi = {10.1093/mnras/stz2794},
archivePrefix = {arXiv},
       eprint = {1905.00868},
 primaryClass = {astro-ph.CO},
       adsurl = {https://ui.adsabs.harvard.edu/abs/2019MNRAS.490.3705S}
}

@ARTICLE{Kay2008,
   author = {{Kay}, S.~T. and {Powell}, L.~C. and {Liddle}, A.~R. and {Thomas}, P.~A.
	},
    title = "{The Sunyaev-Zel'dovich temperature of the intracluster medium}",
  journal = {\mnras},
archivePrefix = "arXiv",
   eprint = {0706.3668},
     year = 2008,
    month = jun,
   volume = 386,
    pages = {2110-2114},
      doi = {10.1111/j.1365-2966.2008.13183.x},
   adsurl = {http://adsabs.harvard.edu/abs/2008MNRAS.386.2110K}
}

@ARTICLE{Fabbri1981,
   author = {{Fabbri}, R.},
    title = "{Spectrum of the Sunyaev-Zel'dovich effect for high electron temperatures}",
  journal = {\apss},
     year = 1981,
    month = jul,
   volume = 77,
    pages = {529-537},
      doi = {10.1007/BF00649478},
   adsurl = {http://adsabs.harvard.edu/abs/1981Ap%26SS..77..529F}
}

@ARTICLE{Hurier2016,
   author = {{Hurier}, G.},
    title = "{High significance detection of the tSZ effect relativistic corrections}",
  journal = {\aap},
archivePrefix = "arXiv",
   eprint = {1701.09020},
     year = 2016,
    month = dec,
   volume = 596,
      eid = {A61},
    pages = {A61},
      doi = {10.1051/0004-6361/201629726},
   adsurl = {http://adsabs.harvard.edu/abs/2016A%26A...596A..61H}
}

@ARTICLE{Erler2017,
   author = {{Erler}, J. and {Basu}, K. and {Chluba}, J. and {Bertoldi}, F.
	},
    title = "{Planck's view on the spectrum of the Sunyaev-Zeldovich effect}",
  journal = {\mnras},
archivePrefix = "arXiv",
   eprint = {1709.01187},
     year = 2018,
    month = may,
   volume = 476,
    pages = {3360-3381},
      doi = {10.1093/mnras/sty327},
   adsurl = {http://adsabs.harvard.edu/abs/2018MNRAS.476.3360E}
}

@ARTICLE{Chluba2012SZpack,
   author = {{Chluba}, J. and {Nagai}, D. and {Sazonov}, S. and {Nelson}, K.
	},
    title = "{A fast and accurate method for computing the Sunyaev-Zel'dovich signal of hot galaxy clusters}",
  journal = {\mnras},
archivePrefix = "arXiv",
   eprint = {1205.5778},
     year = 2012,
    month = oct,
   volume = 426,
    pages = {510-530},
      doi = {10.1111/j.1365-2966.2012.21741.x},
   adsurl = {http://adsabs.harvard.edu/abs/2012MNRAS.426..510C}
}

@ARTICLE{Dai2014,
   author = {{Dai}, L. and {Chluba}, J.},
    title = "{New operator approach to the CMB aberration kernels in harmonic space}",
  journal = {\prd},
archivePrefix = "arXiv",
   eprint = {1403.6117},
     year = 2014,
    month = jun,
   volume = 89,
   number = 12,
      eid = {123504},
    pages = {123504},
      doi = {10.1103/PhysRevD.89.123504},
   adsurl = {http://adsabs.harvard.edu/abs/2014PhRvD..89l3504D}
}

@ARTICLE{Chluba2014mSZII,
   author = {{Chluba}, J. and {Dai}, L.},
    title = "{Multiple scattering Sunyaev-Zeldovich signal - II. Relativistic effects}",
  journal = {\mnras},
archivePrefix = "arXiv",
   eprint = {1309.3274},
 primaryClass = "astro-ph.CO",
     year = 2014,
    month = feb,
   volume = 438,
    pages = {1324-1334},
      doi = {10.1093/mnras/stt2277},
   adsurl = {http://adsabs.harvard.edu/abs/2014MNRAS.438.1324C}
}

@ARTICLE{Chluba2014mSZI,
   author = {{Chluba}, J. and {Dai}, L. and {Kamionkowski}, M.},
    title = "{Multiple scattering Sunyaev-Zeldovich signal - I. Lowest order effect}",
  journal = {\mnras},
archivePrefix = "arXiv",
   eprint = {1308.5969},
 primaryClass = "astro-ph.CO",
     year = 2014,
    month = jan,
   volume = 437,
    pages = {67-76},
      doi = {10.1093/mnras/stt1861},
   adsurl = {http://adsabs.harvard.edu/abs/2014MNRAS.437...67C}
}

@ARTICLE{Refregier2000,
   author = {{Refregier}, A. and {Komatsu}, E. and {Spergel}, D.~N. and {Pen}, U.-L.},
    title = "{Power spectrum of the Sunyaev-Zel'dovich effect}",
  journal = {\prd},
   eprint = {arXiv:astro-ph/9912180},
     year = 2000,
    month = jun,
   volume = 61,
   number = 12,
      eid = {123001},
    pages = {123001},
      doi = {10.1103/PhysRevD.61.123001},
   adsurl = {http://adsabs.harvard.edu/abs/2000PhRvD..61l3001R}
}

@ARTICLE{Chluba2012moments,
   author = {{Chluba}, J. and {Switzer}, E. and {Nelson}, K. and {Nagai}, D.
	},
    title = "{Sunyaev-Zeldovich signal processing and temperature-velocity moment method for individual clusters}",
  journal = {\mnras},
archivePrefix = "arXiv",
   eprint = {1211.3206},
 primaryClass = "astro-ph.CO",
     year = 2013,
    month = apr,
   volume = 430,
    pages = {3054-3069},
      doi = {10.1093/mnras/stt110},
   adsurl = {http://adsabs.harvard.edu/abs/2013MNRAS.430.3054C}
}

@ARTICLE{Challinor2002,
   author = {{Challinor}, A. and {van Leeuwen}, F.},
    title = "{Peculiar velocity effects in high-resolution microwave background experiments}",
  journal = {\prd},
   eprint = {arXiv:astro-ph/0112457},
     year = 2002,
    month = may,
   volume = 65,
   number = 10,
    pages = {103001-+},
      doi = {10.1103/PhysRevD.65.103001},
   adsurl = {http://adsabs.harvard.edu/abs/2002PhRvD..65j3001C}
}

@ARTICLE{Nozawa2006,
   author = {{Nozawa}, S. and et al.},
    title = "{An improved formula for the relativistic corrections to the kinematical Sunyaev-Zeldovich effect for clusters of galaxies}",
  journal = {Nuovo Cimento B Serie},
   eprint = {arXiv:astro-ph/0507466},
     year = 2006,
    month = may,
   volume = 121,
    pages = {487-500},
      doi = {10.1393/ncb/i2005-10223-0},
   adsurl = {http://adsabs.harvard.edu/abs/2006NCimB.121..487N}
}

@ARTICLE{Mroczkowski2012,
   author = {{Mroczkowski}, T. and {Dicker}, S. and {Sayers}, J. and {Reese}, E.~D. and 
	{Mason}, B. and {Czakon}, N. and {Romero}, C. and {Young}, A. and 
	{Devlin}, M. and {Golwala}, S. and {Korngut}, P. and {Sarazin}, C. and 
	{Bock}, J. and {Koch}, P.~M. and {Lin}, K.-Y. and {Molnar}, S.~M. and 
	{Pierpaoli}, E. and {Umetsu}, K. and {Zemcov}, M.},
    title = "{A Multi-wavelength Study of the Sunyaev-Zel'dovich Effect in the Triple-Merger Cluster MACS J0717.5+3745 with MUSTANG and Bolocam}",
  journal = {ArXiv e-prints},
archivePrefix = "arXiv",
   eprint = {1205.0052},
 primaryClass = "astro-ph.CO",
     year = 2012,
    month = apr,
   adsurl = {http://adsabs.harvard.edu/abs/2012arXiv1205.0052M}
}

@ARTICLE{Mroczkowski2019,
       author = {{Mroczkowski}, Tony and {Nagai}, Daisuke and {Basu}, Kaustuv and {Chluba}, Jens and {Sayers}, Jack and {Adam}, R{\'e}mi and {Churazov}, Eugene and {Crites}, Abigail and {Di Mascolo}, Luca and {Eckert}, Dominique and {Macias-Perez}, Juan and {Mayet}, Fr{\'e}d{\'e}ric and {Perotto}, Laurence and {Pointecouteau}, Etienne and {Romero}, Charles and {Ruppin}, Florian and {Scannapieco}, Evan and {ZuHone}, John},
        title = "{Astrophysics with the Spatially and Spectrally Resolved Sunyaev-Zeldovich Effects. A Millimetre/Submillimetre Probe of the Warm and Hot Universe}",
      journal = {SSR},
         year = 2019,
        month = feb,
       volume = {215},
       number = {1},
          eid = {17},
        pages = {17},
          doi = {10.1007/s11214-019-0581-2},
archivePrefix = {arXiv},
       eprint = {1811.02310},
 primaryClass = {astro-ph.CO},
       adsurl = {https://ui.adsabs.harvard.edu/abs/2019SSRv..215...17M}
}

@ARTICLE{Zemcov2012,
   author = {{Zemcov}, M. and {Aguirre}, J. and {Bock}, J. and {Bradford}, C.~M. and 
	{Czakon}, N. and {Glenn}, J. and {Golwala}, S.~R. and {Lupu}, R. and 
	{Maloney}, P. and {Mauskopf}, P. and {Million}, E. and {Murphy}, E.~J. and 
	{Naylor}, B. and {Nguyen}, H. and {Rosenman}, M. and {Sayers}, J. and 
	{Scott}, K.~S. and {Zmuidzinas}, J.},
    title = "{High Spectral Resolution Measurement of the Sunyaev-Zel'dovich Effect Null with Z-Spec}",
  journal = {\apj},
archivePrefix = "arXiv",
   eprint = {1202.0029},
 primaryClass = "astro-ph.CO",
     year = 2012,
    month = apr,
   volume = 749,
      eid = {114},
    pages = {114},
      doi = {10.1088/0004-637X/749/2/114},
   adsurl = {http://adsabs.harvard.edu/abs/2012ApJ...749..114Z}
}

@ARTICLE{Carlstrom2002,
   author = {{Carlstrom}, J.~E. and {Holder}, G.~P. and {Reese}, E.~D.},
    title = "{Cosmology with the Sunyaev-Zel'dovich Effect}",
  journal = {\araa},
   eprint = {arXiv:astro-ph/0208192},
     year = 2002,
   volume = 40,
    pages = {643-680},
      doi = {10.1146/annurev.astro.40.060401.093803},
   adsurl = {http://adsabs.harvard.edu/abs/2002ARA%26A..40..643C}
}

@ARTICLE{Birkinshaw1999,
   author = {{Birkinshaw}, M.},
    title = "{The Sunyaev-Zel'dovich effect}",
  journal = {Phys.~Rep},
   eprint = {arXiv:astro-ph/9808050},
     year = 1999,
    month = mar,
   volume = 310,
    pages = {97-195},
      doi = {10.1016/S0370-1573(98)00080-5},
   adsurl = {http://adsabs.harvard.edu/abs/1999PhR...310...97B}
}

@ARTICLE{Nozawa2005,
   author = {{Nozawa}, S. and {Itoh}, N. and {Kohyama}, Y.},
    title = "{Relativistic corrections to the Sunyaev-Zeldovich effect for clusters of galaxies: effect of the motion of the observer}",
  journal = {\aap},
   eprint = {arXiv:astro-ph/0501114},
     year = 2005,
    month = sep,
   volume = 440,
    pages = {39-44},
      doi = {10.1051/0004-6361:20052923},
   adsurl = {http://adsabs.harvard.edu/abs/2005A%26A...440...39N}
}

@ARTICLE{Hansen2002,
   author = {{Hansen}, S.~H. and {Pastor}, S. and {Semikoz}, D.~V.},
    title = "{First Measurement of Cluster Temperature Using the Thermal Sunyaev-Zel'dovich Effect}",
  journal = {\apjl},
   eprint = {arXiv:astro-ph/0205295},
     year = 2002,
    month = jul,
   volume = 573,
    pages = {L69-L71},
      doi = {10.1086/342094},
   adsurl = {http://adsabs.harvard.edu/abs/2002ApJ...573L..69H}
}

@ARTICLE{Sunyaev1980,
   author = {{Sunyaev}, R.~A. and {Zeldovich}, I.~B.},
    title = "{The velocity of clusters of galaxies relative to the microwave background - The possibility of its measurement}",
  journal = {\mnras},
     year = 1980,
    month = feb,
   volume = 190,
    pages = {413-420},
   adsurl = {http://adsabs.harvard.edu/abs/1980MNRAS.190..413S}
}

@ARTICLE{Wright1979,
   author = {{Wright}, E.~L.},
    title = "{Distortion of the microwave background by a hot intergalactic medium}",
  journal = {\apj},
     year = 1979,
    month = sep,
   volume = 232,
    pages = {348-351},
      doi = {10.1086/157294},
   adsurl = {http://adsabs.harvard.edu/abs/1979ApJ...232..348W}
}

@ARTICLE{Nozawa1998SZ,
   author = {{Nozawa}, S. and {Itoh}, N. and {Kohyama}, Y.},
    title = "{Relativistic Corrections to the Sunyaev-Zeldovich Effect for Clusters of Galaxies. II. Inclusion of Peculiar Velocities}",
  journal = {\apj},
   eprint = {arXiv:astro-ph/9804051},
     year = 1998,
    month = nov,
   volume = 508,
    pages = {17-24},
      doi = {10.1086/306401},
   adsurl = {http://adsabs.harvard.edu/abs/1998ApJ...508...17N}
}

@ARTICLE{Chluba2005b,
   author = {{Chluba}, J. and {H{\"u}tsi}, G. and {Sunyaev}, R.~A.},
    title = "{Clusters of galaxies in the microwave band: Influence of the motion of the Solar System}",
  journal = {\aap},
   eprint = {arXiv:astro-ph/0409058},
     year = 2005,
    month = may,
   volume = 434,
    pages = {811-817},
      doi = {10.1051/0004-6361:20041942},
   adsurl = {http://adsabs.harvard.edu/abs/2005A%26A...434..811C}
}

@ARTICLE{Chluba2011ab,
   author = {{Chluba}, J.},
    title = "{Fast and accurate computation of the aberration kernel for the cosmic microwave background sky}",
  journal = {\mnras},
     year = 2011,
    month = aug,
   volume = 415,
    pages = {3227-3236},
      doi = {10.1111/j.1365-2966.2011.18934.x},
   adsurl = {http://adsabs.harvard.edu/abs/2011MNRAS.415.3227C}
}

@ARTICLE{Challinor1998,
   author = {{Challinor}, A. and {Lasenby}, A.},
    title = "{Relativistic Corrections to the Sunyaev-Zeldovich Effect}",
  journal = {\apj},
   eprint = {arXiv:astro-ph/9711161},
     year = 1998,
    month = may,
   volume = 499,
    pages = {1-+},
      doi = {10.1086/305623},
   adsurl = {http://adsabs.harvard.edu/abs/1998ApJ...499....1C}
}

@ARTICLE{Sazonov1998,
   author = {{Sazonov}, S.~Y. and {Sunyaev}, R.~A.},
    title = "{Cosmic Microwave Background Radiation in the Direction of a Moving Cluster of Galaxies with Hot Gas: Relativistic Corrections}",
  journal = {\apj},
     year = 1998,
    month = nov,
   volume = 508,
    pages = {1-5},
      doi = {10.1086/306406},
   adsurl = {http://adsabs.harvard.edu/abs/1998ApJ...508....1S}
}

@ARTICLE{Zeldovich1969,
   author = {{Zeldovich}, Y.~B. and {Sunyaev}, R.~A.},
    title = "{The Interaction of Matter and Radiation in a Hot-Model Universe}",
  journal = {\apss},
     year = 1969,
    month = jul,
   volume = 4,
    pages = {301-316},
      doi = {10.1007/BF00661821},
   adsurl = {http://adsabs.harvard.edu/abs/1969Ap%26SS...4..301Z}
}

@ARTICLE{Itoh98,  
    author = {{Itoh}, N. and {Kohyama}, Y. and {Nozawa}, S.},
    title = "{Relativistic Corrections to the Sunyaev-Zeldovich Effect for Clusters of Galaxies}",
  journal = {\apj},
   eprint = {arXiv:astro-ph/9712289},
     year = 1998,
    month = jul,
   volume = 502,
    pages = {7-+},
      doi = {10.1086/305876},
   adsurl = {http://adsabs.harvard.edu/abs/1998ApJ...502....7I}
}

@ARTICLE{Sunyaev1970,
   author = {{Sunyaev}, R.~A. and {Zeldovich}, Y.~B.},
    title = "{Small-Scale Fluctuations of Relic Radiation}",
  journal = {\apss},
     year = 1970,
   volume = 7,
    pages = {3-+},
   adsurl = {http://cdsads.u-strasbg.fr/cgi-bin/nph-bib_query?bibcode=1970Ap%26SS...7....3S&db_key=AST}
}

@INPROCEEDINGS{Sehgal2019CMBHD,
       author = {{Sehgal}, Neelima and {Aiola}, Simone and {Akrami}, Yashar and {Basu}, Kaustuv and {Boylan-Kolchin}, Michael and {Bryan}, Sean and {Clesse}, S{\'e}bastien and {Cyr-Racine}, Francis-Yan and {Di Mascolo}, Luca and {Dicker}, Simon and {Essinger-Hileman}, Thomas and {Ferraro}, Simone and {Fuller}, George and {Han}, Dongwon and {Hasselfield}, Matthew and {Holder}, Gil and {Jain}, Bhuvnesh and {Johnson}, Bradley R. and {Johnson}, Matthew and {Klaassen}, Pamela and {Madhavacheril}, Mathew and {Mauskopf}, Philip and {Meerburg}, Daan and {Meyers}, Joel and {Mroczkowski}, Tony and {M{\"u}nchmeyer}, Moritz and {Naess}, Sigurd Kirkevold and {Nagai}, Daisuke and {Namikawa}, Toshiya and {Newburgh}, Laura and {Nguyen}, Nam and {Niemack}, Michael and {Oppenheimer}, Benjamin D. and {Pierpaoli}, Elena and {Schaan}, Emmanuel and {Slosar}, An{\v{z}}e and {Spergel}, David and {Switzer}, Eric and {van Engelen}, Alexander and {Wollack}, Edward},
        title = "{CMB-HD: An Ultra-Deep, High-Resolution Millimeter-Wave Survey Over Half the Sky}",
    booktitle = {Bulletin of the American Astronomical Society},
         year = 2019,
       volume = {51},
        month = sep,
          eid = {6},
        pages = {6},
          doi = {10.48550/arXiv.1906.10134},
archivePrefix = {arXiv},
       eprint = {1906.10134},
 primaryClass = {astro-ph.CO},
       adsurl = {https://ui.adsabs.harvard.edu/abs/2019BAAS...51g...6S}
}

@ARTICLE{ChlubaBO25,
       author = {{Chluba}, Jens and {Ravenni}, Andrea},
        title = "{The Boost Operator: Properties, Computation and Applications}",
      journal = {arXiv e-prints},
         year = 2025,
        month = may,
          eid = {arXiv:2505.02080},
        pages = {arXiv:2505.02080},
          doi = {10.48550/arXiv.2505.02080},
archivePrefix = {arXiv},
       eprint = {2505.02080},
 primaryClass = {astro-ph.CO},
       adsurl = {https://ui.adsabs.harvard.edu/abs/2025arXiv250502080C}
}

@ARTICLE{Lee2024SZpack,
       author = {{Lee}, Elizabeth and {Chluba}, Jens},
        title = "{The SZ effect with anisotropic distributions and high energy electrons}",
      journal = {\jcap},
         year = 2024,
        month = jul,
       volume = {2024},
       number = {7},
          eid = {040},
        pages = {040},
          doi = {10.1088/1475-7516/2024/07/040},
archivePrefix = {arXiv},
       eprint = {2403.18530},
 primaryClass = {astro-ph.HE},
       adsurl = {https://ui.adsabs.harvard.edu/abs/2024JCAP...07..040L}
}

@ARTICLE{Rosenberg2025prSZ,
       author = {{Rosenberg}, Erik and {Chluba}, Jens},
        title = "{Boost operator approach to the relativistic polarized SZ effect}",
      journal = {arXiv e-prints},
         year = 2025,
        month = nov,
          eid = {arXiv:2511.11377},
        pages = {arXiv:2511.11377},
          doi = {10.48550/arXiv.2511.11377},
archivePrefix = {arXiv},
       eprint = {2511.11377},
 primaryClass = {astro-ph.CO},
       adsurl = {https://ui.adsabs.harvard.edu/abs/2025arXiv251111377R}
}

@article{Sayers_2013,
doi = {10.1088/0004-637X/778/1/52},
url = {https://doi.org/10.1088/0004-637X/778/1/52},
year = {2013},
month = {nov},
publisher = {The American Astronomical Society},
volume = {778},
number = {1},
pages = {52},
author = {Sayers, J. and others},
title = {A MEASUREMENT OF THE KINETIC SUNYAEV–ZEL'DOVICH SIGNAL TOWARD MACS J0717.5+3745},
journal = {The Astrophysical Journal}
}

@article{ refId0,
	author = {{Adam} and others},
	title = {Mapping the kinetic Sunyaev-Zel\'{}dovich effect toward  MACS J0717.5+3745 with NIKA },
	DOI= "10.1051/0004-6361/201629182",
	url= "https://doi.org/10.1051/0004-6361/201629182",
	journal = {A\&A},
	year = 2017,
	volume = 598,
	pages = "A115",
}

@ARTICLE{hoey2026derivationkompaneetsequationusing,
doi = {10.1088/1475-7516/2026/06/045},
url = {https://doi.org/10.1088/1475-7516/2026/06/045},
year = {2026},
month = {jun},
publisher = {IOP Publishing},
volume = {2026},
number = {06},
pages = {045},
author = {Hoey, Alex and Long, Jacob and Chluba, Jens},
title = {Derivation of the Kompaneets equation using the boost operator approach},
journal = {Journal of Cosmology and Astroparticle Physics}
}

@ARTICLE{Chluba2026SZ,
       author = {{Chluba}, Jens and {Rosenberg}, Erik},
        title = "{Boost operator approach to the relativistic SZ effect}",
      journal = {\mnras},
         year = 2026,
        month = mar,
       volume = {547},
       number = {1},
          eid = {stag240},
        pages = {stag240},
          doi = {10.1093/mnras/stag240},
archivePrefix = {arXiv},
       eprint = {2508.20659},
 primaryClass = {astro-ph.CO},
       adsurl = {https://ui.adsabs.harvard.edu/abs/2026MNRAS.547ag240C}
}
}

\appendix

\section{Alternative approaches}
\label{app : alt app}
The boost operator approach has been 
applied previously to the case of the kSZ \citep{Chluba2026SZ}, though only to obtain the Thomson collision term for an isotropic lab frame photon distribution, before introducing a more complicated anisotropic electron distribution. This is in contrast to the method we propose, where we consider an additional set of boosts, allowing the average over all directions and electron momenta to be performed in the frame where the electron distribution is isotropic. Here we give a brief overview of the derivation set out by \citep{Chluba2026SZ,Rosenberg2025prSZ} in order to highlight the simplifications presented in our derivation.   

The Thomson collision term for a single velocity scattering event is given by
\bealf{
\label{eq:coll term}
\frac{\text{d}n(\nu, \vgh)}{\text{d}\tau}
&= \left\{\sum_{\ell}Y_{\ell 0}(\vgh)\left[{}^{}\hat{\mathcal{D}}^0_{\ell00}(\nu, \beta)+\frac{1}{10}{}^{}\hat{\mathcal{D}}^0_{\ell20}(\nu, \beta)-\delta_{\ell 0} + \frac{\beta}{\sqrt{3}}\delta_{\ell 1}\right]\right\}n_{00}(\nu).
}
The Doppler operators arise from boosting into and out of the electron rest frame. For the derivation of this collision term, see e.g., \cite{ChlubaBO25}, \cite{Chluba2026SZ}, \cite{hoey2026derivationkompaneetsequationusing}. To obtain the spectral distortion due to the SZ effect, one must both average over all directions $\text{d}\vbh$ and thermally average over all electron momenta. In the lab frame, the electron distribution is anisotropic due to its bulk motion. The distribution function for electrons with bulk velocity $\boldsymbol{\beta}_{\text{p}}$ is given by \citep{Lee2024SZpack}
\bealf{
\label{eq:electron dist}
f_{\text{p}}(\boldsymbol{p})=\frac{(1/\gamma_{\text{p}})\exp{\left(-\frac{\gamma_{\text{p}}\gamma}{\The}\right)}}{4\pi \The K_2(1/\The)}\times\exp{\left(\frac{\boldsymbol{p}_{\text{p}}\cdot\boldsymbol{p}}{\The}\right)},
}
where $\boldsymbol{p}=\boldsymbol{\beta}\gamma$ and $\boldsymbol{p}_{\text{p}}=\boldsymbol{\beta}_{\text{p}}\gamma$ are the dimensionless momenta of the electrons and the cloud respectively. This leads to a spectral distortion given by
\bealf{
\label{eq:Ssz correction rosenberg}
\Delta n_{\textrm{SZ}}(\nu,\vgh)
\approx \tau \, \Sll{}{\textrm{SZ}}(\nu,\The,\vgh,\boldsymbol{\beta}_{\rm p})\,n^{\textrm{Pl}}(\nu),
}
where $\Sll{}{\textrm{SZ}}$ is the SZ scattering operator, which describes how the multipole coefficients of the photon occupation are modified by the scattering process\footnote{We note a minus sign error in the $\beta$ term in Eq.~24 of \citep{Chluba2026SZ} that is corrected in \citep{Rosenberg2025prSZ}}:
\bsub
\bealf{
\label{eq:collect terms 1}
\Sll{}{\textrm{SZ}}(\nu,\The,\vgh,\boldsymbol{\beta}_{\rm p})&=\sum_{\ell=0}^\infty \Sll{}{\ell}(\nu,\The, \beta_{\rm p}) P_\ell (\vgh\cdot\vbh_{\text{p}}),
\\
\label{eq:collect terms 2}
\Sll{}{\ell}(\nu,\The, \beta_{\rm p})
&= \int p^2 f_{\ell}(\gamma,\gamma_{\text{p}})\,\Sll{}{\ell}(\nu, p)\,\text{d}p
,
\\
\label{eq:collect terms 3}
\Sll{}{\ell}(\nu, p)
&= \sqrt{2\ell+1}\left[{}^{}\hat{\mathcal{D}}^0_{\ell00}(\nu, \beta)+\frac{1}{10}{}^{}\hat{\mathcal{D}}^0_{\ell20}(\nu, \beta)-\delta_{\ell 0} + \frac{\beta}{\sqrt{3}}\delta_{\ell 1}\right]
,
\\
\label{eq:collect terms 4}
f_{\ell}(\gamma,\gamma_{\text{p}}) &= \frac{\exp{\left(\frac{-\gamma_{\text{p}}\gamma}{\The}\right)}}{\gamma_{\text{p}}\The K_2(1/\The)}\sqrt{\frac{\pi \, \The}{2p_{\text{p}}p}}I_{\ell+\frac{1}{2}}\left(\frac{p_{\text{p}}p}{\The}\right).
}
\esub
Here, $I_{n}$ are the modified Bessel functions of first kind and $P_\ell$ are the Legendre polynomials. The electron distribution has been decomposed into Legendre moments $f_\ell$ and the velocities $\boldsymbol{\beta}$ and $\boldsymbol{\beta}_{\text{p}}$ are coupled within the operator. To evaluate the spectral distortion, $f_\ell$ must be expanded in orders of $\boldsymbol{\beta}_{\text{p}}$ so that the thermal average [Eq.~\eqref{eq:collect terms 2}] can be performed.
\section{Operator expansions}
\label{operator expansions app}
The thermal SZ operators $\Sll{\textrm{th}}{\ell}$ [see Eq.~\eqref{eq:Sll}] for $\ell=0,1,2,3$, evaluated to $\mathcal{O}(p^8)$, are given by
\bsub
\bealf{
\label{eq:Cloud S0}
\Sll{\textrm{th}}{0}(\nu,p)
&\approx \frac{1}{3}\oDnu p^2+\frac{7}{150}\oDnu(\oDnu-4)p^4+\frac{11}{3150}\oDnu(\oDnu-4)(\oDnu-10)p^6
\nonumber\\
&\qquad
+ \frac{16}{99225}\oDnu(\oDnu-4)(\oDnu-10)(\oDnu-18)p^8
\\
\Sll{\textrm{th}}{1}(\nu,p)
&\approx -1-\frac{2}{15}(1+\oDnu) p^2-\frac{1}{75}(\oDnu-4)(1+2\oDnu)p^4+\frac{1}{7350}(\oDnu-4)(\oDnu-10)(2+17\oDnu)p^6
\nonumber\\
&\qquad
-\frac{1}{198450}(\oDnu-4)(\oDnu-10)(\oDnu-18)(23\oDnu-4)p^8
\\
\Sll{\textrm{th}}{2}(\nu,p)
&\approx -\frac{9}{10}+\frac{1}{30}(\oDnu-6) p^2+\frac{144+5\oDnu(\oDnu-7)}{525}p^4
+\frac{(\oDnu-10)[720+\oDnu(23\oDnu-146)]}{22050}p^6
\nonumber\\
&\qquad
+\frac{2}{99225}(\oDnu-10)(\oDnu-18)[100+\oDnu(3\oDnu-19)]p^8
\\
\Sll{\textrm{th}}{3}(\nu,p)
&\approx -1-\frac{\oDnu-4}{70} p^2-\frac{(\oDnu-4)(\oDnu-12)}{350}p^4-\frac{(\oDnu-4)[3600+\oDnu(23\oDnu-482)]}{66150}p^6
\nonumber\\
&\qquad
-\frac{(\oDnu-4)(\oDnu-18)[1200+\oDnu(7\oDnu-136)]}{297675}p^8.
}
\esub
We highlight that all thermal SZ operators are built from the diffusion operator $\oDnu$, which greatly simplifies the expressions.

The Doppler operators required to derive Eqs.~\eqref{szoperatorsexpanded3rdorder}, \eqref{eq:Ssz expansion} and \eqref{eq:first order T}, to $\mathcal{O}(p_\text{p}^3)$, are given by 
\bsub
\bealf{
\Dbo{}{000}{}&\approx 
1+\oDnu\frac{p_{\rm p}^2}{3}, 
&\Dbo{}{020}{}&\approx \mathcal{O}(p_{\rm p}^4),
\\
\Dbo{}{010}{}&\approx 
-\oDnu\frac{p_{\rm p}^2}{3},
&\Dbo{}{030}{}&\approx 
\mathcal{O}(p_{\rm p}^6),
\\
\Dbo{}{100}{}&\approx 
(\oOnu-1)\frac{p_{\rm p}}{\sqrt{3}}+[15+\oOnu(7-6\oOnu+8\oDnu)]\frac{p_{\rm p}^3}{30\sqrt{3}}, 
&\Dbo{}{120}{}&\approx 
2\oOnu(\oDnu-4)\frac{p_{\rm p}^3}{15\sqrt{3}},
\\
\Dbo{}{110}{}&\approx 
-\oOnu\frac{p_{\rm p}}{\sqrt{3}}+\oOnu(3+2\oOnu-4\oDnu)\frac{p_{\rm p}^3}{10\sqrt{3}},
&\Dbo{}{130}{}&\approx 
\mathcal{O}(p_{\rm p}^5),
\\
\Dbo{}{200}{}&\approx 
(2\oOnu+\oDnu)\frac{p_{\rm p}^2}{3\sqrt{5}}, 
&\Dbo{}{220}{}&\approx 
(\oDnu+4\oOnu)\frac{p_{\rm p}^2}{3\sqrt{5}},
\\
\Dbo{}{210}{}&\approx 
-2(\oDnu+3\oOnu)\frac{p_{\rm p}^2}{3\sqrt{5}},
&\Dbo{}{230}{}&\approx 
\mathcal{O}(p_{\rm p}^4),
\\
\Dbo{}{300}{}&\approx 
\oOnu(-1+3\oOnu+\oDnu)\frac{p_{\rm p}^3}{15\sqrt{7}}, 
&\Dbo{}{320}{}&\approx 
\oOnu(1+5\oOnu+\oDnu)\frac{p_{\rm p}^3}{5\sqrt{7}},
\\
\Dbo{}{310}{}&\approx 
-(1+\oOnu)(\oDnu+3\oOnu)\frac{p_{\rm p}^3}{5\sqrt{7}},
&\Dbo{}{330}{}&\approx 
-\oOnu(2+6\oOnu+\oDnu)\frac{p_{\rm p}^3}{15\sqrt{7}},
}
\esub
which were obtained using {\tt Mathematica}.

\newpage

\section{Closed form derivation}
\label{app: closed form arsinh derivation}
We begin by defining
\bealf{
\label{eq:f(x)}
f(x) = \frac{\cosh{(k\,\text{arsinh}(x))}}{\sqrt{1+x^2}} = \frac{\cosh{(ky)}}{\cosh{(y)}}=F(y),
}
with $y=\text{arsinh}(x)$, such that $\cosh{(y)}=\sqrt{1+x^2}$. We continue by seeking to write a differential equation for $f(x)$, from which we can insert a power series solution. The first and second derivatives of $F(y)$ are given by
\bsub
\bealf{
\label{eq:F'(y)}
F'(y) &= \frac{k\sinh{(ky)}\cosh{(y)}-\cosh{(ky)}\sinh{(y)}}{\cosh^2(y)}, \\
\label{eq:F''(y)}
F''(y) 
&= (k^2-1)\frac{\cosh{(ky)}}{\cosh{y}} - 2\tanh{(y)}\frac{k\sinh{(ky)}\cosh{(y)}-\cosh{(ky)}\sinh{(y)}}{\cosh^2(y)}
\nonumber\\
&= 
(k^2-1)F-2\tanh{(y)}F',
}
\esub
which we can rewrite as a differential equation in $F$,
\bealf{
\label{eq:differential}
F'' +2\tanh{(y)}F' - (k^2-1)F = 0.
}
However, we want to obtain a differential equation in $x$. Since $x=\sinh{(y)}$, it is evident that
\bealf{
\label{eq:dx}
&\frac{\text{d}F}{\text{d}y}=\sqrt{1+x^2}\frac{\text{d}f}{\text{d}x},
&\frac{\text{d}^2F}{\text{d}y^2} = (1+x^2)\frac{\text{d}^2f}{\text{d}x^2} + x \frac{\text{d}f}{\text{d}x}.
}
After inserting the above expressions into Eq.~\eqref{eq:differential}, we have
\bealf{
\label{eq:differential x}
(1+x^2)f'' + 3xf' - (k^2 - 1)f = 0,
}
where $'$ now denotes the derivative with respect to $x$. At this point, we assume a power series solution for $f(x)$. Since we know $f(x)$ is even, we have $f(x)=\sum_{n=0}^\infty a_n x^{2n}$, where $a_n$ are coefficients that we are looking to obtain. Inserting this solution into the differential equation, we have
\bealf{
\label{eq:differential power series}
\sum_{n=1}^\infty 2n(2n-1)a_nx^{2n-2}+ \sum_{n=1}^\infty 2n(2n-1)a_nx^{2n} +  \sum_{n=1}^\infty 6na_nx^{2n}-(k^2-1)\sum_{n=0}^\infty a_n x^{2n}=0.
}
Shifting the index on the first sum ($n\rightarrow n+1$) and equating powers of $x$, we obtain a recursion relation for the coefficients $a_n$,
\bealf{
\label{eq:a_n recursion}
a_{n+1}= \frac{k^2 - (2n+1)}{(2n+2)(2n+1)}a_n,
}
and since we know $a_0=f(0)=1$, we find a general expression for $a_n$,
\bealf{
\label{eq:a_n}
a_n = \sum_{n=0}^\infty \frac{1}{(2n)!}\prod_{m=0}^{n-1}[k^2-(2m+1)^2],
}
and therefore a closed form expansion for $f(x)$,
\bealf{
\label{eq:closed form final}
\frac{\cosh{(k\,\text{arsinh}(x))}}{\sqrt{1+x^2}} = \sum_{n=0}^\infty \frac{1}{(2n)!}\prod_{m=0}^{n-1}[k^2-(2m+1)^2]x^{2n}
}
as needed in the main text.

\section{Additional explicit expressions}
\label{app:additional explicit}
By inspecting coefficients, we find
\bealf{
\label{eq:D0l0_expanded2_dm1}
{}^{-1}\hat{\mathcal{D}}_{0\ell0}
&=
\sum_{n=0}^{\infty}(-1)^{\ell}(2\ell+1)\frac{2^{2n+1}\,p^{2n}}{[2(n+1)]!}\,\frac{(n)_\ell}{(n+2)^{(\ell)}}\,\prod_{m=1}^{n}\left\{\oDnu+2-m(m+1)\right\},
}
where we used the Pochhammer function, $(n)_\ell$ and $(n)^\ell$.
For $\Dbo{-1}{111}{\pm1}$ we find
\bealf{
\label{eq:D111m1_expanded2_dm1}
{}^{-1}\hat{\mathcal{D}}_{111}^{\pm1}
&=
\frac{1}{\oDnu(\oDnu+2)}
\sum_{n=0}^{\infty}\frac{2^{2n+1}\,p^{2n}}{[2(n+1)]!}\,\frac{18(n+1)}{(n+2)(n+3)(2n+3)}\,\prod_{m=0}^{n+1}\left\{\oDnu+2-m(m+1)\right\} \nonumber\\
&=
\sum_{n=0}^{\infty}\frac{2^{2n+1}\,p^{2n}}{[2(n+1)]!}\,\frac{18(n+1)}{(n+2)(n+3)(2n+3)}\,\prod_{m=0}^{n-1}\left\{\oDnu-4-m(m+5)\right\},
}
and for $\Dbo{-1}{121}{\pm1}$ we obtain
\bealf{
\label{eq:D121m1_expanded3_dm1}
{}^{-1}\hat{\mathcal{D}}_{121}^{\pm1}
&=
-\sum_{n=1}^{\infty}\frac{1}{5}\prod_{k=1}^{n-1}\frac{2(2+k)}{k(5+k)(5+2k)}\prod_{m=0}^{n-1}[\oDnu-4-m(m+5)]p^{2n}.
}
Once again by comparing coefficients, we find 
\bealf{
\label{eq:Dlllml_expanded_dm1}
{}^{-1}\hat{\mathcal{D}}_{\ell\ell\ell}^{\pm\ell}
&=
\sum_{n=0}^{\infty}\frac{2^{2n+1}\,p^{2n}}{[2(n+1)]!}\,\frac{\alpha_\ell(n+1)^{(\ell)}}{(n+2)^{(2\ell)}\,2^\ell\,(n+\frac{3}{2})_\ell}\,\prod_{m=0}^{n-1}\left\{\oDnu-\ell(\ell+3)-m(m+2\ell+3)\right\},
}
where $\alpha_{\ell}=18,900,88200,14288400,3457792800$ for $\ell=1,2,3,4,5$.
 A more general combination that may be useful is $\frac{1}{\gamma}{}^a\mathcal{K}_{00}^{0}(\beta)\,{}^b\mathcal{K}_{00}^{0}(\beta)$. We can rewrite this as
\bealf{
\label{eq:doublekernel}
\frac{1}{\gamma}{}^a\mathcal{K}_{00}^{0}(\beta)\,{}^b\mathcal{K}_{00}^{0}(\beta)&=\frac{1}{\gamma} \Bigg[\frac{p_+^{1-a}-p_-^{1-a}}{2p\,(1-a)}\Bigg]\Bigg[\frac{p_+^{1-b}-p_-^{1-b}}{2p\,(1-b)}\Bigg]\\ \nonumber
&
=\frac{1}{2\gamma p^2\,(1-a)(1-b)}\Bigg[\cosh{\Bigg((2-a-b)\,\textrm{arsinh}(p)\Bigg)}-\cosh{\Bigg((a-b)\,\textrm{arsinh}(p)\Bigg)}\Bigg].}
Using the expansion given in Appendix \ref{app: closed form arsinh derivation}, we can then again obtain closed-form expressions which we omit here. We also stress that $a$ and $b$ can generally contain operators. Additional simplifications might be possible by using the hyperbolic function addition theorems, however we  did not explore this any further.

\end{document}